\RequirePackage[l2tabu, orthodox]{nag}
\documentclass[12pt,a4paper]{article}

\usepackage{fixltx2e}

\usepackage[font=small]{caption}
\usepackage[text={450pt,650pt},headheight={14.5pt},centering]{geometry}

\usepackage{lmodern}

\usepackage{graphicx}
\usepackage{booktabs}
\usepackage{subfig}
\usepackage{tikz}
\usetikzlibrary{plotmarks,calc,decorations,decorations.pathmorphing}

\tikzstyle dynkin node=[very thick,shape=circle,draw,inner sep=0pt,minimum size=5mm]
\tikzstyle dynkin line=[very thick]
\tikzstyle inverse line=[gray,line width=1.46pt,line cap=round, dash pattern=on 0pt off 2\pgflinewidth]
\tikzstyle red phase=[red,decoration={snake,amplitude=0.1mm,segment length=1.6mm},decorate]
\tikzstyle blue phase=[blue,decoration={snake,amplitude=0.1mm,segment length=0.9mm},decorate]

\tikzset{%
  hw box/.style={thick,draw,rounded corners,double},
  box/.style={thick,draw,rounded corners},
  Q/.style={thick,rounded corners,red},
  QQ/.style={thick,rounded corners,blue}
}

\pgfdeclarelayer{background}
\pgfdeclarelayer{foreground}
\pgfsetlayers{background,main,foreground}

\usepackage{dsfont}
\usepackage{collref}

\usepackage{amsmath,amssymb}
\usepackage{mathrsfs}
\usepackage{mathtools}
\usepackage{bbm}
\usepackage{slashed}
\usepackage{braket}
\newcommand{\ketbra}[2]{\ket{#1}\!\!\bra{#2}}

\DeclareMathAlphabet{\mathsfit}{\encodingdefault}{\sfdefault}{m}{sl}

\usepackage{hyperref}

\usepackage{xspace}

\numberwithin{equation}{section}

\makeatletter
 \let\old@startsection=\@startsection
 \let\oldl@section=\l@section
 \renewcommand{\@startsection}[6]{\old@startsection{#1}{#2}{#3}{#4}{#5}{#6\mathversion{bold}}}
 \renewcommand{\l@section}[2]{\oldl@section{\mathversion{bold}#1}{#2}}
\makeatother

\DeclareMathOperator{\tr}{tr}

\def\XXint#1#2#3{{\setbox0=\hbox{$#1{#2#3}{\int}$}
    \vcenter{\hbox{$#2#3$}}\kern-.5\wd0}}

\newcommand{\AdS}{\mathrm{AdS}}
\newcommand{\CFT}{\mathrm{CFT}}
\newcommand{\Sphere}{\mathrm{S}}
\newcommand{\Torus}{\mathrm{T}}

\newcommand{\comm}[2]{[#1,#2]}
\newcommand{\acomm}[2]{\{#1,#2\}}

\newcommand{\alg}[1]{\mathrm{#1}}
\newcommand{\grp}[1]{\mathrm{#1}}

\newcommand{\algD}[1]{\alg{d}(2,1;#1)}

\newcommand{\algSL}{\alg{sl}}
\newcommand{\grpSL}{\grp{SL}}

\newcommand{\algSU}{\alg{su}}

\newcommand{\algSO}{\alg{so}}

\newcommand{\algU}{\alg{u}}
\newcommand{\grpU}{\grp{U}}

\newcommand{\algPSU}{\alg{psu}}

\newcommand{\gen}[1]{\mathbf{#1}}

\newcommand{\SymN}{\operatorname{Sym}^{N}}
\newcommand{\SymInfty}{\operatorname{Sym}^{\infty}}

\newcommand{\rep}[1]{\mathbf{#1}}

\newcommand{\superN}{\mathcal{N}}

\newcommand{\ie}{\textit{i.e.}\xspace}
\newcommand{\eg}{\textit{e.g.}\xspace}
\newcommand{\cf}{\textit{cf.}\xspace}

\newcommand{\sL}{\mbox{\tiny L}}
\newcommand{\sR}{\mbox{\tiny R}}

\begin{document}
\thispagestyle{empty}

\begin{center}
\textbf{\Large\mathversion{bold} Protected string spectrum in $\AdS_3 / \CFT_2$ from worldsheet integrability} 
\vspace{2em}

\textrm{\large Marco Baggio${}^1$, Olof Ohlsson Sax${}^2$, Alessandro Sfondrini${}^3$,\\ Bogdan Stefa\'nski, jr.${}^4$ and Alessandro Torrielli${}^5$} 

\vspace{2em}

\vspace{1em}
\begingroup\itshape
1. Instituut voor Theoretische Fysica, KU Leuven, Celestijnenlaan 200D, B-3001 Leuven, Belgium\\[0.2cm]

2. Nordita, Stockholm University and KTH Royal Institute of Technology, Roslagstullsbacken 23, SE-106 91 Stockholm, Sweden\\[0.2cm]

3. Institut f\"ur Theoretische Physik, ETH Z\"urich,\\
Wolfgang-Pauli-Str. 27, 8093 Z\"urich, Switzerland.\\[0.2cm]

4. Centre for Mathematical Science, City, University of London,\\
 Northampton Square, EC1V 0HB London, UK\\[0.2cm]

5. Department of Mathematics, University of Surrey, Guildford, GU2 7XH, UK\par\endgroup

\vspace{1em}

\texttt{marco.baggio@fys.kuleuven.be, olof.ohlsson.sax@nordita.org, sfondria@itp.phys.ethz.ch, Bogdan.Stefanski.1@city.ac.uk, a.torrielli@surrey.ac.uk}


\end{center}

\vspace{6em}

\begin{abstract}\noindent
We derive the protected closed-string spectra of $\AdS_3/\CFT_2$ dual pairs with 16 supercharges at arbitrary values of the string tension and of the three-form fluxes.
These follow immediately from the all-loop Bethe equations for the spectra of the integrable worldsheet theories.
Further, representing the underlying integrable systems as spin~chains, we find that  their dynamics involves length-changing interactions and that protected states correspond to gapless excitations above the Berenstein-Maldacena-Nastase vacuum.
In the case of $\AdS_3\times \Sphere^3\times \Torus^4$ the degeneracies of such operators precisely match those of the dual $\CFT_2$ and the supergravity spectrum. On the other hand,
we find that for $\AdS_3\times \Sphere^3\times \Sphere^3\times \Sphere^1$
there are fewer protected states than previous supergravity calculations had suggested. In particular,  protected states have the same $\alg{su}(2)$ charge with respect to the two three-spheres.
\end{abstract}

\newpage

\tableofcontents

\section{Introduction}
\label{sec:introduction}

Recent years have seen significant progress in the application of integrability techniques to the planar sector of the $\AdS_3/\CFT_2$ correspondence. Most of these advances have stemmed from the study of superstrings on $\AdS_3\times\Sphere^3\times\Torus^4$ and $\AdS_3\times\Sphere^3\times\Sphere^3\times\Sphere^1$ backgrounds, which preserve sixteen supercharges and can be supported by a mixture of Ramond-Ramond (RR) and Neveu-Schwarz-Neveu-Schwarz (NSNS) fluxes. In particular, it has been shown that the string non-linear sigma models (NLSM) are classically integrable for these backgrounds~\cite{Babichenko:2009dk,Sundin:2012gc,Cagnazzo:2012se}.%
\footnote{See also ref.~\cite{David:2008yk} for some earlier progress in $\AdS_3/\CFT_2$ integrability.}
 An efficient way to check whether integrability persists at the quantum level is to study the S~matrix that scatters the excitations on the string worldsheet, following the ideas pioneered by Zamolodchikov  and Zamolodchikov for two-dimensional relativistic integrable QFTs~\cite{Zamolodchikov:1978xm}. Indeed, for the cases of $\AdS_3\times\Sphere^3\times\Torus^4$ and $\AdS_3\times\Sphere^3\times\Sphere^3\times\Sphere^1$ supported by arbitrary three-form fluxes, the two-body S~matrix can be fixed by symmetry considerations up to ``dressing factors'' and turns out to automatically satisfy the Yang-Baxter equation~\cite{Borsato:2012ud, Borsato:2013qpa,Borsato:2014exa,Borsato:2014hja,Lloyd:2014bsa, Borsato:2015mma}.%
\footnote{See also ref.~\cite{Sfondrini:2014via} for a review of these constructions.}
The dressing factors are not fixed by symmetry, rather they satisfy crossing equations. So far they have been found only for the case of $\AdS_3\times\Sphere^3\times\Torus^4$ supported by pure-RR fluxes~\cite{Borsato:2013hoa, Borsato:2016xns}.
Equipped with an integrable S~matrix, one may write down the Bethe equations that predict the asymptotic spectrum%
\footnote{%
  This description is valid up to finite-size ``wrapping'' effects~\cite{Ambjorn:2005wa, Abbott:2015pps}.%
} %
of closed strings in these backgrounds; this construction has been recently completed for pure-RR $\AdS_3\times\Sphere^3\times\Torus^4$~\cite{Borsato:2016kbm,Borsato:2016xns}, and preliminary results are known for $\AdS_3\times\Sphere^3\times\Sphere^3\times\Sphere^1$ too~\cite{OhlssonSax:2011ms,Borsato:2012ss}.

The integrability construction interpolates between the perturbative string NLSM regime and the ``perturbative CFT'' regime. For this purpose, we introduce a coupling constant $h$ so that the large-coupling limit $h\gg1$ corresponds to large string tension%
\footnote{%
The precise relation between the coupling $h$ and the string tension is at present only known perturbatively~\cite{Beccaria:2012kb,Sundin:2014sfa}. In the case of integrable $\AdS_4/\CFT_3$ backgrounds an analogous function $h$ was determined exactly by comparing integrability and localisation results~\cite{Gromov:2014eha}.%
}
(and hence a perturbative description on the worldsheet) while $h\ll1$ corresponds to the 't Hooft coupling being small.
The large-$h$ expansion of the integrability construction has been matched against a number of perturbative string NLSM computations in the near-plane-wave, near-flat-space and semi-classical regimes, see \eg~\cite{Rughoonauth:2012qd,Beccaria:2012kb, Abbott:2012dd,Abbott:2013mpa, Sundin:2013ypa, Roiban:2014cia, Sundin:2014ema, Sundin:2015uva, Sundin:2016gqe}.
On the other hand, the small-$h$ limit is more subtle. While it is known that the $\AdS_3\times\Sphere^3\times\Torus^4$ background lives in the moduli space of the a semi-direct product CFT $(\SymN( \Torus^4))\ltimes \Torus^4$~\cite{Maldacena:1997re,Larsen:1999uk}, it is not clear how the symmetric-product orbifold point is related to the $h\to0$ limit of the integrability description nor how to find integrability there~\cite{Pakman:2009mi}. The first evidence of integrability on the $\CFT_2$ side
has instead been found in the IR limit of the Higgs branch of the two-dimensional super-Yang-Mills theory with matter~\cite{Sax:2014mea} and deserves to be explored further. The scenario is even more mysterious for $\AdS_3\times\Sphere^3\times\Sphere^3\times\Sphere^1$, where the identification of what the dual $\CFT_2$ might be remains a challenge~\cite{Gukov:2004ym}.%
\footnote{See however \cite{Tong:2014yna} for a recent proposal.}

In this paper we use integrability to determine the protected closed-string spectra of the $\AdS_3/\CFT_2$ models. We begin with the case of pure-RR $\AdS_3\times \Sphere^3\times \Torus^4$ for which we derive in detail such protected states from the all-loop Bethe equations, elaborating on the results announced in~\cite{Borsato:2016kbm}.  Our findings are valid for generic values of the string tension in the planar theory. The degeneracy we find matches precisely the spectrum of protected operators~\cite{deBoer:1998kjm} and the (modified) elliptic genus of the symmetric-product orbifold~CFT and supergravity~\cite{deBoer:1998us,Maldacena:1999bp}. Let us stress that the protected states' Bethe roots are particularly simple: they carry no worldsheet momentum. As such, determining
the protected spectrum from the Bethe equations is a remarkably straightforward exercise. Furthermore, it is also easy to show that finite-size ``wrapping'' corrections to the Bethe ansatz \emph{exactly cancel} for these protected states.
In a small way, this simplicity indicates the power of integrability in solving the spectral problem.

Based on these results, it becomes clear that protected states can be found for more general $\AdS_3$ integrable backgrounds by classifying zero-momentum Bethe roots. In particular, we find that the spectrum of protected operators for the pure-RR $\AdS_3\times\Sphere^3\times\Sphere^3\times\Sphere^1$ theory consists only of states that have equal $\algSU(2)$ charges $J_{+}=J_{-}$ with respect to the two three-spheres; once again, this result is valid for generic values of the string tension in the planar theory.
We compare our findings with the results available in the literature. At the supergravity point, it has been proposed~\cite{deBoer:1999gea} that protected multiplets with $J_{+}\neq J_{-}$ should exist, yielding a much larger degeneracy. 
On the other hand, at generic points of the moduli space, it is expected~\cite{deBoer:1999gea,Gukov:2004fh} that such multiplets should not be protected.%
\footnote{%
This follows from observing that the Lie-superalgebra shortening condition receives non-linear corrections in the full large $\mathcal{N}=(4,4)$ superconformal algebra precisely when $J_{+}\neq J_{-}$.
}
 Our integrability-based findings give an explicit confirmation of this expectation, though this does not address whether additional degenaricies might appear at the supergravity point. The analysis of ref.~\cite{deBoer:1999gea} is not conclusive as it relies on the assumption that all Ka\l{}u\.za-Klein (KK) modes sit in short representations and thus might overcount the BPS states. A simple yet effective way of checking if protected KK modes with $J_{+}\neq J_{-}$ exist is to study point-like string solutions for the $\AdS_3\times\Sphere^3\times\Sphere^3\times\Sphere^1$ background~\cite{Babichenko:2009dk}, see also~\cite{Lloyd:2013wza,Abbott:2014rca}, which are in one-to-one correspondence with its bosonic supersymmetric ground states. We find that, even in the large-tension limit, the only BPS multiplets are the ones with $J_{+}= J_{-}$. In addition, we show that our integrability analysis of protected states holds also for mixed RR- and NSNS-flux $\AdS_3$ backgrounds.
It is interesting to note that the restriction on the allowed angular momenta $J_{+}= J_{-}$ for $\AdS_3\times\Sphere^3\times\Sphere^3\times\Sphere^1$ was also found for BPS giant-graviton D-strings~\cite{Prinsloo:2014dha}. This suggests that this condition may well apply to the full non-perturbative protected spectrum of the theory.

One of the central lessons of holographic integrability in higher dimensions~\cite{Minahan:2002ve}, see also~\cite{Arutyunov:2009ga,Beisert:2010jr} for a review, has been the appearance of an integrable spin-chain in the small-$h$ regime.
Such a description is very useful for enumerating the states of the theory, and in particular for classifying the protected operators.
As in the higher-dimensional cases, all the pure-RR $\AdS_3$ backgrounds exhibit a lattice-like dispersion relation
\begin{equation}
\label{eq:RRdispersion}
E(p)=\sqrt{m^2+4\,h^2\,\sin^2(\tfrac{p}{2})}\,,
\end{equation} 
where $m$ is the mass of a world-sheet excitation and $p$ its momentum, suggesting that the theory should also have a representation in terms of discretised degrees of freedom. Motivated by this expectation, we construct a spin chain at $h\ll1$ and describe how the protected states we found correspond to inserting gapless excitations at zero momentum above the Berenstein-Maldacena-Nastase (BMN)~\cite{Berenstein:2002jq} vacuum. 
This description is particularly useful for enumerating the states of the theory and computing supersymmetric indices.

The paper is structured as follows. In section~\ref{sec:CFT} we start by briefly reviewing the structure of protected multiplets in $\mathcal{N}=(4,4)$ SCFTs. In section~\ref{sec:BA} we recapitulate the structure of the Bethe equations for pure-RR $\AdS_3\times\Sphere^3\times\Torus^4$ strings at $h\ll1$. Building on these, in section~\ref{sec:spinchain} we construct the weakly-coupled spin chain. It is then straightforward to derive, in section~\ref{sec:fermioniczeromodes}, the spectrum of protected operators and to match that with the results of section~\ref{sec:CFT}. We then describe in section~\ref{sec:s1} how to apply the same techniques to the case of pure-RR $\AdS_3\times\Sphere^3\times\Sphere^3\times\Sphere^1$, derive the spectrum of protected states and corroborate it by a classical string-theory calculation. In section~\ref{sec:mixed} we argue on general grounds that our results are unchanged in the presence of a mixture of RR and NSNS three-form fluxes. Finally, in section~\ref{sec:wrapping} we argue that our results, which where derived from the Bethe equations, remain valid even when wrapping corrections are taken into account. 
We conclude in section~\ref{sec:conclusions}, and relegate some technical details to four appendices.

\subsection*{Note added}
While we were in the final stages of preparing our manuscript, we received~\cite{Eberhardt:2017fsi} where the protected spectrum on $\AdS_3\times\Sphere^3\times\Sphere^3\times\Sphere^1$ is also investigated both at the Wess-Zumino-Witten point by CFT techniques and at the supergravity one through a direct analysis of the KK spectrum. The results perfectly agree with our integrability findings for this background.

\section{BPS states in \texorpdfstring{$\mathcal{N}=(4,4)$}{N=(4,4)} theories}
\label{sec:CFT}

In this section we review the properties of chiral states in theories with \emph{small} and \emph{large} $\mathcal{N}=(4,4)$ superconformal symmetry. The global small $\mathcal{N}=(4,4)$ algebra consists of two copies of the $\algPSU(1,1|2)$ algebra, acting on the left- and right-moving sectors respectively. As reviewed in appendix \ref{sec:psu112}, representations are characterized by the (left and right) conformal dimension $(D\sL,D\sR)$ and the R-symmetry quantum numbers $(J\sL,J\sR)$ of the superconformal primaries. The unitarity bound for the left copy of the algebra in these conventions is
\begin{equation}
\label{eq:bpsconditionSmall}
  D\sL \geq J\sL,
\end{equation}
and similarly for the right one. Multiplets that saturate this bound are called (left) chiral primary multiplets. They contain $2J+1$ superconformal primaries differing in their $\gen{J}$ eigenvalue $j$. The highest-weight state with $j=J$ is annihilated by the supercharges $\gen{Q}_1$ and $\dot{\gen{Q}}^2$.\footnote{The lowest weight state $j=-J$ is annihilated by $\gen{Q}_2$ and $\dot{\gen{Q}}^1$, while generic superconformal primaries with $j \neq \pm J$ are not annihilated by any combination of the supercharges. Instead, they satisfy nullness conditions, in that certain combinations of their superconformal descendants vanish.} As a consequence, chiral primary multiplets are shorter than generic multiplets. Multiplets that saturate the unitarity bound in both the left and right sector are 1/2-BPS, those that saturate it only in one sector are 1/4-BPS. More details on the structure of short and long representations of the $\algPSU(1,1|2)$ algebra are presented in appendix \ref{sec:psu112}.

We now discuss short representations of the large $\mathcal{N}=(4,4)$ algebra. Its global part is given by two copies of $\algD{\alpha}$. For each copy, the R~symmetry is $\algSU(2) \oplus \algSU(2)$, so that representations are labelled by the conformal dimension $D$ and the two $\algSU(2)$ spins $(J_+,J_-)$. The unitarity bound then reads\footnote{In the literature, the parameter $\alpha$ is sometimes denoted by $\gamma$, typically when denoting the superconformal algebra by $\mathcal{A}_{\gamma}$.}
\begin{equation}
\label{eq:bpsconditionLarge}
  D \geq \alpha J_+ + (1-\alpha)J_- .
\end{equation}
The highest-weight states of representations that saturate this bound are annihilated by one supercharge, so such multiplets are shorter than generic ones, but preserve less supersymmetry compared to their small $\mathcal{N} = (4,4)$ counterparts. We then have 1/4-BPS and 1/8-BPS multiplets, depending on whether they saturate the unitarity bound on both left and right sectors or only on one sector. The BPS bound for the full superconformal algebra (as opposed to its global part) differs from \eqref{eq:bpsconditionLarge} when $J_+ \neq J_-$~\cite{Gunaydin:1988re}. However, we will find in section \ref{sec:s1} that the only BPS states for $\AdS_3 \times \Sphere^3 \times \Sphere^3 \times \Sphere^1$ satisfy $J_+ = J_-$, so we will not discuss this point further.

\subsection{Poincar\'e polynomial}

In theories with small $\mathcal{N}=(4,4)$ supersymmetry, the spectrum of 1/2-BPS states can be usefully encoded in the \emph{Poincar\'e polynomial}
\begin{equation}
  \label{eq:poincare-polynomial}
  P_{t,\bar{t}} = \mathrm{Tr}\,t^{2\gen{J}\sL} \bar{t}^{2\gen{J}\sR} ,
\end{equation}
where the trace is taken over the space of highest-weight states of the chiral primary multiplets. This polynomial has finite degree, since the spectrum of chiral primaries is bounded from above \cite{Lerche:1989uy}
\begin{equation}
  h \leq \frac{c}{6} .
\end{equation}
In the case of a sigma model with target space $\mathcal{M}$, it can be shown that the Poincar\'e polynomial is given by \cite{Witten:1981nf}
\begin{equation}
  P_{t,\bar{t}} = \sum_{p,q} h^{p,q} t^p \bar{t}^q,
\end{equation}
where $h^{p,q}$ are the Betti numbers of the target space $\mathcal{M}$. In this paper, we are especially interested in the case $\mathcal{M}=\Torus^4$, whose Betti numbers can be read off from its Hodge diamond
\begin{equation}
  \begin{tikzpicture}[baseline=-0.5ex]
    \node at ( 0.0,+1.0) {$1$};
    \node at (-0.5,+0.5) {$2$};
    \node at (+0.5,+0.5) {$2$};
    \node at (-1.0, 0.0) {$1$};
    \node at ( 0.0, 0.0) {$4$};
    \node at (+1.0, 0.0) {$1$};
    \node at (-0.5,-0.5) {$2$};
    \node at (+0.5,-0.5) {$2$};
    \node at ( 0.0,-1.0) {$1$};
  \end{tikzpicture}
\end{equation}
The Poincar\'e polynomial of the symmetric orbifold $\SymN(\mathcal{M})$ can also be expressed in terms of the Betti numbers of $\mathcal{M}$~\cite{Gottsche:1993perverse,deBoer:1998kjm}:
\begin{equation}
  \label{eq:ppsym}
  \sum_{N} Q^N P_{t,\bar t}(\SymN(\mathcal{M})) = \prod_{m=1}^\infty \prod_{p,q} \bigl( 1+(-1)^{p+q+1}Q^m t^{p+\frac{d}{2}(m-1)} \bar{t}^{q+\frac{d}{2}(m-1)} \bigr)^{(-1)^{p+q+1} h^{p,q}}  ,
\end{equation}
where $d$ is the dimension of $\mathcal{M}$.

The large-$N$ behaviour of $P_{t,\bar t}(\SymN(\mathcal{M}))$ can be extracted using the relation~\cite{deBoer:1998kjm}
\begin{equation}
  P_{t,\bar{t}}(\SymInfty(\mathcal{M})) = \lim_{Q \to 1} (1-Q) \sum_{N=0}^{\infty} Q^N P_{t,\bar{t}}(\SymN(\mathcal{M})) .
\end{equation}
In the case of $\mathcal{M}=\Torus^4$, this gives
\begin{equation}
  \label{eq:ppt4}
  P_{t,\bar{t}}(\SymInfty(\Torus^4)) =\frac{
    (1 + t)^2 (1 +\bar{t})^2
    }{
    (1 - t\, \bar{t})^5
  } \prod_{J=2}^{\infty} \frac{
    (1 + t^{J} \, \bar{t}^{J-1})^4 (1 + t^{J-1} \, \bar{t}^{J})^4
  }{
    (1 - t^{J-2} \, \bar{t}^{J}) (1 - t^{J} \, \bar{t}^{J-2}) (1 - t^{J} \, \bar{t}^{J})^6
  } .
\end{equation}
This exhibits the expected structure of a freely generated partition function: bosonic generators with $\algSU(2)\sL \oplus \algSU(2)\sR$ charge $(J\sL,J\sR)$ contribute a factor of the form
\begin{equation}
  1 + t^{2 J\sL} \bar{t}^{2 J\sR} + t^{4 J\sL} \bar{t}^{4 J\sR} + \ldots = \frac{1}{(1 - t^{2 J\sL} \bar{t}^{2 J\sR})} ,
\end{equation}
while fermionic ones give rise to
\begin{equation}
  1 + t^{2 J\sL} \bar{t}^{2 J\sR}.
\end{equation}
In particular, we can read off from~\eqref{eq:ppt4} the following generators: for $J \geq 2$
\begin{itemize}
  \item 6 bosonic generators with charges $(\tfrac{J}{2}, \tfrac{J}{2})$,
  \item 4+4 fermionic generators with charges $(\tfrac{J-1}{2}, \tfrac{J}{2})$ and $(\tfrac{J}{2}, \tfrac{J-1}{2})$,
  \item 1+1 bosonic generators with charges $(\tfrac{J}{2}-1, \tfrac{J}{2})$ and $(\tfrac{J}{2}, \tfrac{J}{2}-1)$,
\end{itemize}
and additionally, there are
\begin{itemize}
  \item 2+2 fermionic generators with charges $(\tfrac{1}{2}, 0)$ and $(0, \tfrac{1}{2})$
  \item 5 bosonic generators with charges $(\tfrac{1}{2}, \tfrac{1}{2})$ .
\end{itemize}
In section \ref{sec:fermioniczeromodes}, we will derive this exact spectrum of protected states using the all-loop integrable Bethe Equations and show how such degeneracies appear from a spin-chain perspective.

\subsection{(Modified) elliptic genus}
\label{sec:modelliptic}
A more refined quantity that is constant over the moduli space is the \emph{elliptic genus}. This is defined in the NS sector as\footnote{\label{ftn:egramond}This quantity is usually defined in the Ramond sector of the Hilbert space $$ \mathrm{Tr}_{RR} (-1)^F q^{2 (\gen{D}\sL-c/24)}\bar{q}^{2 (\gen{D}\sR-c/24)} y^{2 \gen{J}\sL} ,$$ where it only receives contributions from states of the form $|\textrm{anything}\rangle\sL \otimes |\textrm{ground state}\rangle\sR$. Using spectral flow, it can be related to the quantity in \eqref{eq:ellipticg}~\cite{deBoer:1998us}.
}
\begin{equation}
  \label{eq:ellipticg}
  \mathcal{E}(q,y) = \mathrm{Tr}_{|\textrm{anything}\rangle\sL \otimes |\textrm{chiral primary}\rangle\sR} (-1)^F q^{2 \gen{D}\sL} y^{2 \gen{J}\sL} .
\end{equation}
This quantity vanishes for theories with large $\mathcal{N}=(4,4)$ supersymmetry~\cite{Maldacena:1999bp}. The chiral algebra of the $\SymN(\Torus^4)$ theory, due to the additional $\algU(1)^4$ symmetry associated to translations along the four directions of the torus, is actually a degenerate large $\mathcal{N}=(4,4)$ algebra, so that its elliptic genus vanishes. Therefore, we consider the \emph{modified elliptic genus}~\cite{Maldacena:1999bp}, defined in the NS sector as
\begin{equation}
  \mathcal{E}_2(q,y) = \mathrm{Tr}_{|\textrm{anything}\rangle\sL \otimes |\textrm{chiral primary}\rangle\sR} (-1)^F q^{2 \gen{D}\sL} y^{2 \gen{J}\sL} (2 \gen{J}\sR)^2 .
\end{equation}
In analogy to the Poincar\'e polynomial \eqref{eq:ppsym}, it is useful to introduce the generating function of elliptic genera for the symmetric orbifold:
\begin{equation}
  \tilde{\mathcal{E}}_2(Q,q,y) = \sum_N Q^N \mathcal{E}_2(\SymN(\Torus^4))	 .
\end{equation}
Unlike the Poincar\'e polynomial, the elliptic genus diverges in the large $N$ limit, so the comparison with the spin-chain computation is more subtle. The same issue was encountered in the computation of the elliptic genus in supergravity \cite{deBoer:1998us}, where a suitable notion of degree was introduced to take into account the stringy exclusion principle \cite{Maldacena:1998bw}, see also appendix~\ref{app:elliptic-genus}. In this case, the only terms that can be meaningfully compared are those that correspond to states with dimension $D < N/4$ \cite{Maldacena:1999bp}
\begin{equation}
  \label{eq:leadingeg}
  \tilde{\mathcal{E}}_2(Q,q,y) = \sum_N 2 N Q^N + \ldots .
\end{equation}
We will reproduce this result in the spin-chain picture after introducing an appropriate notion of degree.

\section{The \texorpdfstring{$\AdS_3 \times \Sphere^3 \times \Torus^4$}{AdS3 x S3 x T4} Bethe equations}
\label{sec:BA}

\emph{Bethe equations}~\cite{Bethe:1931hc} are a set of polynomial equations whose solutions---the \emph{Bethe roots}---determine the spectrum of an associated integrable system. In the simplest cases the Bethe roots are in one-to-one correspondence with the momenta of particles or magnons and the Bethe equations are immediately recognisable as quantisation conditions for the momenta, accounting for periodic boundary conditions and for phase shifts due to scattering of the magnons. The energy of a given state is determined through the dispersion relation much like in a free theory. In theories with extra global symmetries, like the ones we consider here, the \emph{momentum-carrying} roots are supplemented by \emph{auxiliary} roots. The latter do not directly affect the dispersion relation, but are necessary to reproduce states with different Noether charges.

The full set of Bethe equations for pure-RR $\AdS_3\times\Sphere^3\times\Torus^4$ strings was found in~\cite{Borsato:2016kbm,Borsato:2016xns}. 
It is convenient to work with the \emph{weak-coupling} Bethe equations~\cite{Borsato:2016xns}, which are valid at small but non-vanishing~$h$. Besides simplifying our formulae, this will allow us to explicitly enumerate the states in a spin-chain language, which is more natural at weak coupling, and still reproduce the generic spectrum of protected states.
The weak-coupling Bethe equations are reproduced in equations~\eqref{eq:be-1}--\eqref{eq:be-3m} below. There are in total nine types of Bethe roots, three momentum-carrying and six auxiliary. For the excitations charged under $\algPSU(1,1|2)_{\sL}$ we have momentum-carrying roots $\{u_k\}_{k=1,\dots M_2}$ and auxiliary roots $\{v_{1,k}\}_{k=1,\dots M_1}$ and $\{v_{3,k}\}_{k=1,\dots M_3}$. Similarly for $\algPSU(1,1|2)_{\sR}$ we write $v_{\bar{1},k}, \bar{u}_k, v_{\bar{3},k}$ and $M_{\bar{1}},M_{\bar{2}},M_{\bar{3}}$. 
Finally, the massless sector is described by the momentum-carrying roots $z_k^\pm$ and the auxiliary roots $r_{1,k}$ and $r_{3,k}$, with excitation numbers $N_0$, $N_1$ and $N_3$.%
\footnote{The massless momentum-carrying roots are conveniently parametrised using two complex conjugate parameters $z_k^+$ and $z_k^-$ which satisfy $z_k^+ z_k^- = 1$.} The momentum-carrying Bethe roots are directly related to the momentum of the corresponding excitations through the relations
\begin{equation}
  \frac{u_j + \tfrac{i}{2}}{u_j - \tfrac{i}{2}} = e^{ip_j} , \quad
  \frac{\bar{u}_j + \tfrac{i}{2}}{\bar{u}_j - \tfrac{i}{2}} = e^{ip_j} , \quad
  \frac{z_j^+}{z_j^-} = e^{ip_j},
\label{eq:defn-of-momenta}
\end{equation}
where in the three equations above for simplicity of notation we indicate with the same $p_j$ the momentum of a given left-massive, right-massive and massless excitation, respectively.
Note that for real momenta $u_j,\bar{u}_j\in\mathbb{R}$ and $|z^\pm_j|=1$.
The energy of a given state can be deduced via the weak-coupling limit of the dispersion relation~\eqref{eq:RRdispersion}, which in terms of the Bethe roots is given below in equation~\eqref{eq:anomalousdim} .

The massive left-moving Bethe roots satisfy the $\algPSU(1,1|2)_{\sL}$ Bethe equations
\begin{align}
  1 &= \prod_{j=1}^{M_2} \frac{v_{1,k} - u_j - \frac{i}{2}}{v_{1,k} - u_j + \frac{i}{2}} , \label{eq:be-1}
  \\
  \left(\frac{u_k + \frac{i}{2}}{u_k - \frac{i}{2}}\right)^{\!\!L-N_0+N_1+N_3}
    &= \prod_{\substack{j = 1\\j \neq k}}^{M_2} \frac{u_k - u_j + i}{u_k - u_j - i}
  \prod_{j=1}^{M_1} \frac{u_k - v_{1,j} - \frac{i}{2}}{u_k - v_{1,j} + \frac{i}{2}}
  \prod_{j=1}^{M_3} \frac{u_k - v_{3,j} - \frac{i}{2}}{u_k - v_{3,j} + \frac{i}{2}} , \label{eq:be-2}
  \\
  1 &= \prod_{j=1}^{M_2} \frac{v_{3,k} - u_j - \frac{i}{2}}{v_{3,k} - u_j + \frac{i}{2}} , \label{eq:be-3}
\end{align}
while the massive right-moving roots satisfy the $\algPSU(1,1|2)_{\sR}$ equations
\begin{align}
  1 &= \prod_{j=1}^{M_{\bar{2}}} \frac{v_{\bar{1},k} - \bar{u}_j + \frac{i}{2}}{v_{\bar{1},k} - \bar{u}_j - \frac{i}{2}} , \label{eq:be-1b}
  \\
  \left(\frac{\bar{u}_k + \frac{i}{2}}{\bar{u}_k - \frac{i}{2}}\right)^{\!\!L}
    &= \prod_{\substack{j = 1\\j \neq k}}^{M_{\bar{2}}} \frac{\bar{u}_k - \bar{u}_j - i}{\bar{u}_k - \bar{u}_j + i}
  \prod_{j=1}^{M_{\bar{1}}} \frac{\bar{u}_k - v_{\bar{1},j} + \frac{i}{2}}{\bar{u}_k - v_{\bar{1},j} - \frac{i}{2}}
  \prod_{j=1}^{M_{\bar{3}}} \frac{\bar{u}_k - v_{\bar{3},j} + \frac{i}{2}}{\bar{u}_k - v_{\bar{3},j} - \frac{i}{2}} , \label{eq:be-2b}
  \\
  1 &= \prod_{j=1}^{M_{\bar{2}}} \frac{v_{\bar{3},k} - \bar{u}_j + \frac{i}{2}}{v_{\bar{3},k} - \bar{u}_j - \frac{i}{2}} . \label{eq:be-3b}
\end{align}
These two sets of equations differ from each other in two ways. Firstly, the equations for the left movers are written in an $\algSU(2)$ grading (\ie~feature the Heisenberg $\algSU(2)$ S~matrix) while the equations for the right movers are written in an $\algSL(2)$ grading.%
\footnote{At weak coupling the two sets of equations are completely decoupled and the grading of each can be individually changed. However, at higher orders in the coupling the two sets of equations are coupled by extra interaction factors and the two gradings need to be chosen in a consistent way. Hence, we find it convenient to use different gradings also at weak coupling.}
Secondly, the lengths appearing in the driving terms on the left hand sides of equations~\eqref{eq:be-2} and~\eqref{eq:be-2b} are different whenever the excitation numbers~$N_j$ corresponding to the massless Bethe roots are non-zero. As we will see below this feature is essential for obtaining a spin-chain interpretation of the full set of Bethe equations including the massless modes.

The massless Bethe roots satisfy the equations
\begin{align}
  1 &= \prod_{j=1}^{N_0} \frac{r_{1,k} - z_j^+}{r_{1,k} - z_j^-}   \prod_{j=1}^{M_2} \frac{u_j+\tfrac{i}{2}}{u_j-\tfrac{i}{2}} , \label{eq:be-1m}
  \\
  \left(\frac{z_k^+}{z_k^-}\right)^{\!\!L}
  &= \prod_{\substack{j = 1\\j \neq k}}^{N_0} \frac{z_k^+ - z_j^-}{z_k^- - z_j^+} ( \sigma_{kj}^{\circ\circ} )^2
  \prod_{j=1}^{M_2} \frac{u_j - \tfrac{i}{2}}{u_j +\tfrac{i}{2}}
  \prod_{j=1}^{N_1} \frac{z_k^- - r_{1,j}}{z_k^+ - r_{1,j}}
  \prod_{j=1}^{N_3} \frac{z_k^- - r_{3,j}}{z_k^+ - r_{3,j}}
  , \label{eq:be-0}
  \\
  1 &= \prod_{j=1}^{N_0} \frac{r_{3,k} - z_j^+}{r_{3,k} - z_j^-}   \prod_{j=1}^{M_2} \frac{u_j+\tfrac{i}{2}}{u_j-\tfrac{i}{2}} , \label{eq:be-3m}
\end{align}
where $\sigma_{kj}^{\circ\circ}$ is the massless-massless dressing factor~\cite{Borsato:2016kbm,Borsato:2016xns}. Note that the massless roots couple to the \emph{total momentum} carried by the massive left-moving roots $u_k$.

Finally, the three types of momentum-carrying roots are coupled through the level-matching constraint
\begin{equation}
  1 = 
  \prod_{k=1}^{M_2} \frac{u_k + \tfrac{i}{2}}{u_k - \tfrac{i}{2}}
  \prod_{k=1}^{M_{\bar{2}}} \frac{\bar{u}_k + \tfrac{i}{2}}{\bar{u}_k - \tfrac{i}{2}}
  \prod_{k=1}^{N_0} \frac{z_k^+}{z_k^-} .
\end{equation}
and enter the dispersion relation as
\begin{equation}
\label{eq:anomalousdim}
\delta D = h^2 \sum_{j=1}^{M_2} \frac{2}{1+4u_j}+h^2 \sum_{j=1}^{M_{\bar{2}}} \frac{2}{1+4\bar{u}_j}+i\, h\sum_{j=1}^{N_0} \left(\frac{1}{z^+_j}-\frac{1}{z^-_j}\right)\,,
\end{equation}
where $\delta D$ denotes the anomalous dimension.
It is interesting to note that in the weak-coupling limit, massless modes contribute one order earlier than massive ones, as expected from~\eqref{eq:RRdispersion}.

The Cartan charges corresponding to a solution to the Bethe equations are given in terms of the excitation numbers by
\begin{equation}
  \label{eq:charges}
  \begin{aligned}
    D &= \phantom{D_{\sL}} \mathllap{D_{\sL}} + \mathrlap{D_{\sR}} \phantom{D_{\sL}}
    = L + M_{\bar{2}} + \tfrac{1}{2} \bigl( M_1 + M_3 + N_1 + N_3 - M_{\bar{1}} - M_{\bar{3}} - N_0 \bigr) + \delta D , \\
    J &= \phantom{D_{\sL}} \mathllap{J_{\sL}} + \mathrlap{J_{\sR}} \phantom{D_{\sL}}
    = L - M_2 + \tfrac{1}{2} \bigl( M_1 + M_3 + N_1 + N_3 - M_{\bar{1}} - M_{\bar{3}} - N_0 \bigr) , \\
    S &= \phantom{D_{\sL}} \mathllap{D_{\sL}} - \mathrlap{D_{\sR}} \phantom{D_{\sL}}
    = \phantom{L} - M_{\bar{2}} + \tfrac{1}{2} \bigl( M_1 + M_3 + N_1 + N_3 + M_{\bar{1}} + M_{\bar{3}} - N_0 \bigr) , \\
    K &= \phantom{D_{\sL}} \mathllap{J_{\sL}} - \mathrlap{J_{\sR}} \phantom{D_{\sL}}
    = \phantom{L} - M_2 + \tfrac{1}{2} \bigl( M_1 + M_3 + N_1 + N_3 + M_{\bar{1}} + M_{\bar{3}} - N_0 \bigr) .
  \end{aligned}
\end{equation}
As discussed in~\cite{Borsato:2016xns} (see also~\cite{Borsato:2014exa,Borsato:2014hja}) massless excitations are charged under an additional $\algSU(2)_{\circ}$ that commutes with $\algPSU(1,1|2)_{\sL} \oplus \algPSU(1,1|2)_{\sR}$. This symmetry is implicit in the Bethe equations and yields a $2^{N_0}$-fold degeneracy of the spectrum. We can easily account for it by arbitrarily assigning an eigenvalue $\pm 1/2$ to each massless momentum-carrying root.

The global $\algPSU(1,1|2)_{\sL} \oplus \algPSU(1,1|2)_{\sR}$ symmetry manifests itself through the possibility of adding extra massive Bethe roots at infinity without affecting the Bethe equations or the dispersion relation~\eqref{eq:anomalousdim}. Additionally, as discussed in~\cite{Borsato:2016xns} we can add an arbitrary number of bosonic massless excitations at zero momentum ($z^\pm_j=1$);\footnote{The apparent symmetry when adding a massless root at
$|z^{\pm}_j|=\infty$ is spurious, as the massless rapidity is constrained to $|z^{\pm}_j|=1$.%
}
 this too is a symmetry corresponding to shifts along $\Torus^4$ in target space.
Adding a \emph{fermionic} massless zero mode to the Bethe equations \emph{is not a symmetry unless} $M_2=M_{\bar{2}}=0$. As we will see in section~\ref{sec:fermioniczeromodes}, in that case the fermionic zero modes precisely generate the protected states expected from the discussion of section~\ref{sec:CFT}.

\section{Spin-chain interpretation}
\label{sec:spinchain}

Let us now see how the structure of the weak coupling Bethe equations discussed above can arise from a spin-chain picture of local operators. Such a formulation is very helpful for enumerating the solutions of the Bethe equations, and hence the states of the theory. In particular, as we will see below, it will give us a concrete description of the protected operators at weak coupling. 
The structure we find here follows directly from the Bethe equations by introducing an appropriate set of fields that to describe local operators. This notation is purposefully reminiscent of that introduced in~\cite{Sax:2014mea} in the context of integrability on the Higgs branch of the dual CFT.

\subsection{Massive excitations}

To start with we will consider the massive sector by setting the excitation numbers $N_0$, $N_1$ and $N_3$ of the massless Bethe roots to zero. This case was already discussed in~\cite{OhlssonSax:2011ms,Borsato:2013qpa}. The massive spin-chain is homogenous with the sites transforming in the 1/2-BPS $(\tfrac{1}{2};\tfrac{1}{2}) \otimes (\tfrac{1}{2};\tfrac{1}{2})$ representation of $\algPSU(1,1|2)_{\sL} \oplus \algPSU(1,1|2)_{\sR}$. Since this representation will be important in the following discussion let us review its construction.

The representation $(\tfrac{1}{2};\tfrac{1}{2})$ of $\algPSU(1,1|2)$ consists of two bosons $\phi^{\pm}$ transforming as a doublet under the $\algSU(2)$ R~symmetry of $\algPSU(1,1|2)$ and two fermions $\psi^{\pm}$ transforming as a doublet under the outer $\algSU(2)_{\bullet}$ automorphism of $\algPSU(1,1|2)$. Additionally there are derivatives $\partial$ that generate the $\algSU(1,1)$ descendants of the bosons and fermions. The bosons have $\algSU(1,1)$ weight $1/2$ while the fermions have weight $1$. Explicit expressions for the action of the generators on the states can be found in, \eg, appendix~B of~\cite{OhlssonSax:2011ms}.

The symmetry of $\AdS_3 \times \Sphere^3 \times \Torus^4$ includes two copies of $\algPSU(1,1|2)$ and the sites of the massive spin chain transform under the same $(\tfrac{1}{2};\tfrac{1}{2})$ representation under both copies. However, only the diagonal outer automorphism $\algSU(2)_{\bullet}$ is a symmetry.%
\footnote{This $\algSU(2)_{\bullet}$  can be seen as arising from an $\alpha\to0$ contraction of $\algD{\alpha}\sL\oplus\algD{\alpha}\sR$, or geometrically as part of the $\algSO(4)$ symmetry in the $\Torus^4$ directions in the NLSM target space; it is sometimes referred to as ``custodial'' $\algSU(2)$, see for example~\cite{Gukov:2004ym}.
}
 The fields appearing at the site of this spin chain are listed in table~\ref{tab:massive-fields}.
\begin{table}
  \centering
  \begin{tabular}{cr@{$\,\otimes\,$}lcccccc}
    \toprule
    & \multicolumn{2}{c}{} & $D$ & $S$ & $\algSU(2)_{\sL}$ & $\algSU(2)_{\sR}$ & $\algSU(2)_{\bullet}$ & $\algSU(2)_{\circ}$\\
    \midrule
    $\phi^{\alpha\dot{\alpha}}$ & $\phi^{\dot{\alpha}}$ & $\phi^{\alpha}$ & $1$ & $\phantom{+}0$ & $\rep{2}$ & $\rep{2}$ & $\rep{1}$ & $\rep{1}$ \\
    $\psi_{\sL}^{\alpha\dot{a}}$ & $\psi^{\dot{a}}$ & $\phi^{\alpha}$ & $\frac{3}{2}$ & $+\tfrac{1}{2}$ & $\rep{1}$ & $\rep{2}$ & $\rep{2}$ & $\rep{1}$ \\[2pt]
    $\psi_{\sR}^{\dot{\alpha}\dot{a}}$ & $\phi^{\dot{\alpha}}$ & $\psi^{\dot{a}}$ & $\tfrac{3}{2}$ & $-\tfrac{1}{2}$ & $\rep{2}$ & $\rep{1}$ & $\rep{2}$ & $\rep{1}$ \\
    $F^{\dot{a}\dot{b}}$ & $\psi^{\dot{a}}$ & $\psi^{\dot{b}}$ & $2$ & $\phantom{+}0$ & $\rep{1}$ & $\rep{1}$ & $\rep{3} + \rep{1}$ & $\rep{1}$ \\
    \midrule
    $\nabla_{\sL}$ & $\partial$ & $1$ & $1$ & $+1$ & $\rep{1}$ & $\rep{1}$ & $\rep{1}$ & $\rep{1}$ \\
    $\nabla_{\sR}$ & $1$ & $\partial$ & $1$ & $-1$ & $\rep{1}$ & $\rep{1}$ & $\rep{1}$ & $\rep{1}$ \\
    \bottomrule
  \end{tabular}

  \caption{Primary fields appearing at the sites of the massive spin chain. The first column gives the various fields, and the second column gives their decomposition under $\algPSU(1,1|2)_{\sL} \oplus \algPSU(1,1|2)_{\sR}$. The rest of the table shows how the fields transform under the bosonic subalgebra as well as under the two extra symmetries $\algSU(2)_{\bullet}$ and $\algSU(2)_{\circ}$. The derivatives in the last two lines of the table are not fields but are included to indicate the quantum numbers carried by $\algSU(1,1)$ descendants of the fields. Note that the field $F^{\dot{a}\dot{b}}$ can be decomposed into a triplet and a singlet under $\algSU(2)_{\bullet}$ by writing $F^{\dot{a}\dot{b}} = D^{\dot{a}\dot{b}} + \epsilon^{\dot{a}\dot{b}} F$, where $D^{\dot{a}\dot{b}}$ is symmetric. \label{tab:massive-fields}}
\end{table}

The Bethe equations are constructed with respect to a ground state of the form\footnote{%
  The states we write can be thought of as cyclic ``single-trace'' operators. However, since we are often interested in more general operators that do not satisfy the level matching constraint we do not write out an explicit trace.%
}%
\begin{equation}
\label{eq:BMNvacuumT4}
  \ket{ (\phi^{++})^L } .
\end{equation}
This state carries charges $D_{\sL} = J_{\sL} = D_{\sR} = J_{\sR} = L/2$ and hence satisfies the 1/2-BPS conditions~\eqref{eq:bpsconditionSmall}.

To obtain an excited state we replace the $\phi^{++}$ sitting at one or more of the sites with one of the other fields in table~\ref{tab:massive-fields}. This can be interpreted as acting on the field at those sites with lowering operators of $\algPSU(1,1|2)_{\sL} \oplus \algPSU(1,1|2)_{\sR}$. 

A generic state will contain several excitations and transform as a long representation of $\algPSU(1,1|2)_{\sL} \oplus \algPSU(1,1|2)_{\sR}$. However, let us consider the case where we have only a single left-moving excitation, so that the state takes the form\footnote{%
\label{foot:magnon}
  When we write a sum over permutations as in~\eqref{eq:state-su2L} this is meant to indicate linear combination of states where the excitation appears at different sites. In general we consider a state of definite spin chain momentum $p$, such as $ \sum_{n=0}^L e^{ipn} \ket{ (\phi^{++})^n \phi^{-+} (\phi^{++})^{L-n-1} }$.
  However, as the exact form of the state is not important for this discussion, we only indicate the field content.
}%
\begin{equation}
  \label{eq:state-su2L}
  \ket{ (\phi^{++})^{L-1} \phi^{-+} } + \text{permutations} .
\end{equation}
If the excitation has non-zero momentum this is a highest weight state with charges\footnote{%
  We use the notation $(D_{\sL}, J_{\sL} ; D_{\sR} , J_{\sR})_{J_{\bullet}}$ to denote the eigenvalues of a state under the Cartan elements $D_{\sL}$, $J_{\sL}$, $D_{\sR}$, $J_{\sR}$ of $\algPSU(1,1|2) \oplus \algPSU(1,1|2) \oplus \algSU(2)_{\bullet}$.%
} %
$( \tfrac{L}{2} , \tfrac{L}{2} - 1 ; \tfrac{L}{2} , \tfrac{L}{2} )_0$ up to anomalous corrections of order~$h^2$ given by eq.~\eqref{eq:anomalousdim}. Since the state has $D_{\sL} > J_{\sL}$ but $D_{\sR} = J_{\sR}$ it should transform in a long $\algPSU(1,1|2)_{\sL}$ representation, but in a short representation of $\algPSU(1,1|2)_{\sR}$.
This would imply that the charge $D_{\sR} - J_{\sR}$ is protected, while $D_{\sL} - J_{\sL}$ can receive continuous corrections. However, this cannot be since the charges $S = D_{\sL} - D_{\sR}$ and $K = J_{\sL} - J_{\sR}$ are compact and hence have quantised eigenvalues. 
In fact, as long as the momentum of the excitation is non-vanishing, we expect from the Bethe equations that $D_{\sL}+D_{\sR}$ is corrected by $\delta D>0$, \cf~\eqref{eq:anomalousdim}--\eqref{eq:charges}.
The resolution of this puzzle is that several such 1/4-BPS multiplets join to form long multiplets.

For instance, in the case we analysed, the representation with charges $(\tfrac{L}{2},\tfrac{L}{2} - 1 ; \tfrac{L}{2},\tfrac{L}{2})_0$ is short under $\algPSU(1,1|2)_{\sR}$ and hence has a highest weight state that is annihilated by $\gen{S}_{\sR}^2$ and $\dot{\gen{S}}_{\sR\,1}$, as well as by the raising operators. However, this representation can join up with three more 1/4-BPS representations with highest weight states carrying charges
\begin{equation}\label{eq:massive-L-multiplet-splitting}
  (\tfrac{L}{2},\tfrac{L}{2} - 1 ;  \tfrac{L-1}{2},\tfrac{L-1}{2})_{+\tfrac{1}{2}} , \quad
  (\tfrac{L}{2},\tfrac{L}{2} - 1 ; \tfrac{L-1}{2},\tfrac{L-1}{2})_{-\tfrac{1}{2}} , \quad
  (\tfrac{L}{2},\tfrac{L}{2} - 1 ; \tfrac{L}{2}-1,\tfrac{L}{2}-1)_0 .
\end{equation}
The combined representation can then receive an anomalous dimension $\delta D$ in which case we obtain a single representation with charges
\begin{equation}
  (\tfrac{L + \delta D}{2},\tfrac{L}{2} - 1 ; \tfrac{L + \delta D}{2},\tfrac{L}{2})_0.	
\end{equation}
This deformed representation is long under both $\algPSU(1,1|2)_{\sL}$ and $\algPSU(1,1|2)_{\sR}$.

Using the field content of table~\ref{tab:massive-fields} we can construct the multiplets~\eqref{eq:massive-L-multiplet-splitting} explicitly. Their highest-weight states are
\begin{equation}
\label{eq:massive-L-hws}
\ket{(\phi^{++})^{L-2}\psi^{++}_{\sL}},\qquad
\ket{(\phi^{++})^{L-2}\psi^{+-}_{\sL}},\qquad
\ket{(\phi^{++})^{L-3}\nabla_{\sL}\phi^{++}},
\end{equation}
respectively. As all these states must be part of a long multiplet of $\algPSU(1,1|2)_{\sR}$, they must be related to the action of some right-moving supercharges; namely, in this case, by $\dot{\gen{Q}}_{\sR}^{1}$ and $\gen{Q}_{\sR 2}$. We see here that these supercharges \emph{change the length} of the spin-chain ground-state, as proposed in~\cite{Borsato:2013qpa}. Clearly we could derive a similar picture starting from a right-moving excitation in equation~\eqref{eq:state-su2L}, see appendix~\ref{app:massive-joining} where we also further detail the length-changing action of the supercharges on such multiplets. Furthermore, in Appendix~\ref{app:dynamic-sc} we determine how $\algPSU(1,1|2)_{\sL}\oplus \algPSU(1,1|2)_{\sR}$ generators act on the massive spin-chain for the first few $h$-orders in a more general sub-sector. These length-changing effects are then seen to follow from the closure of the algebra much like in $\AdS_5/\CFT_4$~\cite{Beisert:2003ys}.

\subsection{Massless excitations}
\label{sec:massless-excitations}

Having understood the structure of the massive spin chain, let us now turn to the massless excitations. To start with we consider a simple configuration consisting of one left-moving and one right-moving massive excitation, $M_2 = M_{\bar{2}} = 1$. In the absence of any other Bethe roots these massive roots satisfy the free momentum quantisation conditions
\begin{equation}
  \Bigl( \frac{u + \tfrac{i}{2}}{u - \tfrac{i}{2}} \Bigr)^{L} = 1, \qquad
  \Bigl( \frac{\bar{u} + \tfrac{i}{2}}{\bar{u} - \tfrac{i}{2}} \Bigr)^{L} = 1,
\end{equation}
where $u\equiv u_{j}\text{ with }j=1$, and similarly for $\bar{u}$.
If we now additionally turn on one massless momentum-carrying root by setting $N_0 = 1$ we find that the above equations are modified to
\begin{equation}
  \Bigl( \frac{u + \tfrac{i}{2}}{u - \tfrac{i}{2}} \Bigr)^{L-1} = 1, \qquad
  \Bigl( \frac{\bar{u} + \tfrac{i}{2}}{\bar{u} - \tfrac{i}{2}} \Bigr)^{L} = 1. 
\end{equation}
The new equations still describe two free massive excitations. However, the effective length of the spin chain along which the excitations propagate is different in the left and
right sectors. It is natural to interpret this as the presence of a \emph{chiral} site in the spin chain: at the site where the massless excitation sits the representation $(\tfrac{1}{2};\tfrac{1}{2}) \otimes (\tfrac{1}{2};\tfrac{1}{2})$ has been replaced by $\tilde{\rep{1}}^{a} \otimes (\tfrac{1}{2};\tfrac{1}{2})$, where $\tilde{\rep{1}}^a$ is a singlet under $\algPSU(1,1|2)_{\sL}$.\footnote{As discussed above there is also an additional $\algSU(2)_{\circ}$ symmetry acting only on the massless modes. We take this into account by letting the $\algPSU(1,1|2)$ singlet transform as a doublet (denoted by the index $a$) under this extra symmetry.} From the structure of the world-sheet excitations studied in~\cite{Borsato:2014exa,Borsato:2014hja} we know that the massless highest weight excitation is fermionic. Hence we have included a tilde on the singlet to indicate that it has an odd grading. 

When we add the massless root, we also get an additional equation from~\eqref{eq:be-0}, describing the quantisation of the momentum of the massless excitation. In the simple case of $N_0 = 1$ and $M_2, M_{\bar{2}} \geq 0$ this equation reads
\begin{equation}
\label{eq:1masslessbetheeq}
  \Bigl( \frac{z^+}{z^-} \Bigr)^L = \prod_{j=1}^{M_{2}}\frac{u_j + \tfrac{i}{2}}{u_j - \tfrac{i}{2}} .
\end{equation}
We see that the massless excitation behaves almost like a free excitation propagating on a chain of length $L$, except it feels an extra twist which depends on the total momentum of the massive left-moving excitations, \cf equation~\eqref{eq:defn-of-momenta}.

Let us now consider an even simpler system consisting of only a single massless excitation inserted above a ferromagnetic ground state of length $L$. In order to simplify the description of such states we label the extra fields appearing in the chiral spin chain as in table~\ref{tab:massless-fields}. The state with a single massless excitation then takes the form
\begin{equation}
  \label{eq:single-massless-magnon}
  \ket{ (\phi^{++})^{L-1} \chi_{\sR}^{+a} } + \text{permutations,}
\end{equation}
and carries $\algPSU(1,1|2)_{\sL} \oplus \algPSU(1,1|2)_{\sR}$ charges
\begin{equation}
  ( \tfrac{L-1}{2} , \tfrac{L-1}{2} ; \tfrac{L}{2} , \tfrac{L}{2} )_0 .
\end{equation}
At zero coupling this satisfies the 1/2-BPS conditions $D_{\sL} = J_{\sL}$ and $D_{\sR} = J_{\sR}$. Hence, in the free theory it transforms in a short representation which is annihilated by the creation operators $\gen{Q}_{\sL\,2}$, $\dot{\gen{Q}}_{\sL}^1$, $\gen{S}_{\sR}^2$ and $\dot{\gen{S}}_{\sR\,1}$.
Should its dimension to be protected even when the coupling constant $h$ is non-zero? This cannot be the case in general: firstly it would result in a glut of protected operators~\cite{Sax:2012jv}. Secondly, when the excitation has non-zero momentum and $h>0$, the dispersion relation in equation~\eqref{eq:anomalousdim} shows that the state should receive an anomalous dimension $\delta D$. The mechanism for the dimension of such states to receive corrections comes once again through multiplet joining and length changing.
\begin{table}
  \centering
  \begin{tabular}{cr@{$\,\otimes\,$}lcccccc}
    \toprule
    & \multicolumn{2}{c}{} & $D$ & $S$ & $\algSU(2)_{\sL}$ & $\algSU(2)_{\sR}$ & $\algSU(2)_{\bullet}$ & $\algSU(2)_{\circ}$\\
    \midrule
    $\chi_{\sL}^{\dot{\alpha}a}$ & $\phi^{\dot{\alpha}}$ & $\tilde{\rep{1}}^{a}$ & $\tfrac{1}{2}$ & $+\tfrac{1}{2}$ & $\rep{2}$ & $\rep{1}$ & $\rep{1}$ & $\rep{2}$ \\
    $\nabla_{\sL} T^{a\dot{a}}$ & $\psi^{\dot{a}}$ & $\tilde{\rep{1}}^{a}$ & $1$ & $+1$ & $\rep{1}$ & $\rep{1}$ & $\rep{2}$ & $\rep{2}$ \\
    \midrule
    $\chi_{\sR}^{\alpha a}$ & $\tilde{\rep{1}}^{a}$ & $\phi^{\alpha} $ & $\tfrac{1}{2}$ & $-\tfrac{1}{2}$ & $\rep{1}$ & $\rep{2}$ & $\rep{1}$ & $\rep{2}$ \\
    $\nabla_{\sR} T^{a\dot{a}}$ & $\tilde{\rep{1}}^{a}$ & $\psi^{\dot{a}}$ & $1$ & $-1$ & $\rep{1}$ & $\rep{1}$ & $\rep{2}$ & $\rep{2}$ \\
    \midrule
    $\nabla_{\sL}$ & $\partial$ & $1$ & $1$ & $+1$ & $\rep{1}$ & $\rep{1}$ & $\rep{1}$ & $\rep{1}$ \\
    $\nabla_{\sR}$ & $1$ & $\partial$ & $1$ & $-1$ & $\rep{1}$ & $\rep{1}$ & $\rep{1}$ & $\rep{1}$ \\
    \bottomrule
  \end{tabular}

  \caption{The massless fields decomposed as representations of $\algPSU(1,1|2)_{\sL} \oplus \algPSU(1,1|2)_{\sR}$. The symbol $\tilde{\rep{1}}^{a}$ describes an $\algPSU(1,1|2)_{\sL} \oplus \algPSU(1,1|2)_{\sR}$ singlet with fermionic grading which transforms as a doublet under $\algSU(2)_{\circ}$. \label{tab:massless-fields}}
\end{table}

In order to construct the additional states that are needed to complete a long representation we need one additional ingredient: a bosonic field $T^{a\dot{a}}$ that transforms as a singlet under $\algPSU(1,1|2)_{\sL} \oplus \algPSU(1,1|2)_{\sR}$ and as a bi-spinor under $\algSU(2)_{\bullet} \oplus \algSU(2)_{\circ}$. This field, which emerges naturally in the near-BMN analysis of the string spectrum~\cite{Borsato:2014exa}, represents a spin-chain site that transforms trivially in both the left and the right sector. When we allow for states that include this massless bosonic excitation we can construct a long representation as described in appendix~\ref{app:massless-joining}.

If the massless excitation in~\eqref{eq:single-massless-magnon} has vanishing momentum, its anomalous dimension vanishes even for non-zero coupling. In fact, it is easy to see that this is the case even in the all-loop dispersion relation~\eqref{eq:RRdispersion}. These states are solutions of the Bethe equations that correspond to 1/2-BPS multiplets at arbitrary values of the string tension. We will see in the next section that they precisely reproduce the spectrum of protected closed-string states that we expect from supergravity and from the dual $\CFT_2$.

\section{Fermionic zero modes and protected states}
\label{sec:fermioniczeromodes}

As we have just seen, it is interesting to consider massless excitations with zero momentum on top of the BMN ground state. The resulting states solve the all-loop Bethe equations, and have anomalous dimension $\delta D=0$. As such, they are groundstates, degenerate with the ferromagnetic BMN vacuum.
Since there are four different massless fermionic excitations ($\chi_{\sR}^{+\pm}$ and $\chi_{\sL}^{+\pm}$) we have a total of four such \emph{fermionic zero modes}. 

The Bethe equations ensure that these 1/2-BPS multiplets exist for arbitrary values of the coupling $h$~\cite{Borsato:2016kbm}. It is particularly transparent to construct their highest-weight states in the  spin-chain description that emerges at small but non-vanishing~$h$.
 The ferromagnetic ground state of length $L$ is
\begin{equation}
  \ket{ (\phi^{++})^L } 
\label{eq:bmn-vac}
\end{equation}
and in the Bethe equations its excitation numbers are all set to zero. Turning on a single massless momentum-carrying root ($N_0=1$) we get one of the two states
\begin{equation}
  \ket{ (\phi^{++})^{L-1} \chi_{\sR}^{+\pm} } +\text{symmetric~permutations} ,
\end{equation}
depending on which $\algSU(2)_{\circ}$ spin we assign to the root. Since the excitation has vanishing momentum, the above states are \emph{completely symmetric}, \cf~footnote~\ref{foot:magnon}.
  The corresponding Bethe root is  $z^{\pm} = 1$ and its Bethe equation~\eqref{eq:be-0} is trivially satisfied with both sides equal to one.
Similarly we can get a left-moving fermion
\begin{equation}
  \ket{ (\phi^{++})^{L} \chi_{\sL}^{+\pm} } +\text{symmetric~permutations} ,
\end{equation}
by turning on $N_0 = N_1 = N_3 = 1$. The momentum-carrying root still sits at $z^{\pm} = 1$, as do the two auxiliary roots, $r_1 = r_3 = 1$.

\begin{table}
  \centering
  \begin{tabular}{lcccccc}
    \toprule
    State & $N_0$ & $N_1$ & $N_3$ & $J_{\sL}$ & $J_{\sR}$ & $J_{\circ}$ \\
    \midrule
    $(\phi^{++})^L$ & $0$ & $0$ & $0$ & $\frac{L}{2}$ & $\frac{L}{2}$ & $0$ \\
    \midrule
    $(\phi^{++})^{L-1} \chi_{\sR}^{+\pm}$ & $1$ & $0$ & $0$ & $\frac{L-1}{2}$ & $\frac{L}{2}$ & $\pm\frac{1}{2}$ \\[4pt]
    $(\phi^{++})^{L\phantom{-1}} \chi_{\sL}^{+\pm}$ & $1$ & $1$ & $1$ & $\frac{L+1}{2}$ & $\frac{L}{2}$ & $\pm\frac{1}{2}$ \\
    \midrule
    $(\phi^{++})^{L-2} \chi_{\sR}^{++} \chi_{\sR}^{+-}$ & $2$ & $0$ & $0$ & $\frac{L-2}{2}$ & $\frac{L}{2}$ & $0$ \\[4pt]
    $(\phi^{++})^{L-1} \chi_{\sR}^{+\pm} \chi_{\sL}^{+\pm}$ & $2$ & $1$ & $1$ & $\frac{L}{2}$ & $\frac{L}{2}$ & $\pm 1$ \\[4pt]
    $(\phi^{++})^{L-1} \chi_{\sR}^{+\pm} \chi_{\sL}^{+\mp}$ & $2$ & $1$ & $1$ & $\frac{L}{2}$ & $\frac{L}{2}$ & $0$ \\[4pt]
    $(\phi^{++})^{L\phantom{-1}} \chi_{\sL}^{++} \chi_{\sL}^{+-}$ & $2$ & $1$ & $1$ & $\frac{L+2}{2}$ & $\frac{L}{2}$ & $0$ \\
    \midrule
    $(\phi^{++})^{L-2} \chi_{\sR}^{++} \chi_{\sR}^{+-} \chi_{\sL}^{+\pm}$ & $3$ & $1$ & $1$ & $\frac{L-1}{2}$ & $\frac{L}{2}$ & $\pm\frac{1}{2}$ \\[4pt]    
    $(\phi^{++})^{L-1} \chi_{\sR}^{+\pm} \chi_{\sL}^{++} \chi_{\sL}^{+-}$ & $3$ & $1$ & $1$ & $\frac{L+1}{2}$ & $\frac{L}{2}$ & $\pm\frac{1}{2}$ \\
    \midrule
    $(\phi^{++})^{L-2} \chi_{\sR}^{++} \chi_{\sR}^{+-} \chi_{\sL}^{++} \chi_{\sL}^{+-}$ & $4$ & $2$ & $2$ & $\frac{L}{2}$ & $\frac{L}{2}$ & $0$ \\
    \bottomrule
  \end{tabular}
  \caption{The sixteen 1/2-BPS states obtained from a spin-chain ground state of length $L$ by inserting the four fermionic zero modes, and the corresponding excitation numbers and charges under $\algSU(2)_{\sL}$, $\algSU(2)_{\sR}$ and $\algSU(2)_{\circ}$. \label{tab:spin-chain-BPS-states}}
\end{table}

Note that only the fermions $\chi_{\sR}^{+\pm}$ and $\chi_{\sL}^{+\pm}$ give a new state with the same energy as the corresponding ground state. If we were to insert the fermion $\chi_{\sL}^{-\pm}$ the resulting state would no longer satisfy the 1/2-BPS condition, because the fermion would give a positive contribution to the energy.\footnote{%
  From the Bethe equation point of view, such a state can be interpreted as a multi-excitation state, as discussed at the end of section~\ref{sec:massless-excitations}.%
} %
Hence the next states we can obtain contain two fermionic excitations. For example, to consider a state containing two right-moving massless fermions, 
\begin{equation}
  \ket{ (\phi^{++})^{L-2} \chi_{\sR}^{++} \chi_{\sR}^{+-} } +\text{symmetric~permutations} ,
\end{equation}
we set $N_0 = 2$, with both particles having zero energy, \textit{i.e.}\ roots sitting at $z^{\pm} = +1$ or $z^{\pm} = -1$. Due to fermion anti-symmetry the two excitations have to have opposite $\algSU(2)_{\circ}$ spin.%
\footnote{%
In terms of the Bethe equations this exclusion principle is encoded in the usual requirement that regular solutions must have distinct Bethe roots. This can be seen explicitly by constructing the Bethe wave-function.}
Continuing in this fashion we obtain the sixteen states shown in table~\ref{tab:spin-chain-BPS-states}.

In the grading we use it is natural from the spin-chain perspective to group the BPS states according to their $\algSU(2)_{\sR}$ charge. However, since the states sit in different irreducible representations of $\algPSU(1,1|2)_{\sL} \oplus \algPSU(1,1|2)_{\sR}$ this grouping is completely arbitrary, and a more transparent picture of the BPS states emerges if we instead reorganise the states according to the number of fields $\phi^{++}$ appearing in the operators. The resulting set of states is shown in table~\ref{tab:spin-chain-BPS-states-J}. 
\begin{table}
  \centering
  \begin{tabular}{lccccccc}
    \toprule
    State & $L$ & $J_{\sL}$ & $J_{\sR}$ & $\algSU(2)_{\circ}$ \\
    \midrule
    $(\phi^{++})^J$ & $J$ & $\frac{J}{2}$ & $\frac{J}{2}$ & $\rep{1}$ \\
    \midrule
    $(\phi^{++})^J \chi_{\sR}^{+\pm}$ & $J+1$ & $\frac{J}{2}$ & $\frac{J+1}{2}$ & $\rep{2}$ \\[4pt]
    $(\phi^{++})^J \chi_{\sL}^{+\pm}$ & $J$ & $\frac{J+1}{2}$ & $\frac{J}{2}$ & $\rep{2}$ \\
    \midrule
    $(\phi^{++})^J \chi_{\sR}^{++} \chi_{\sR}^{+-}$ & $J+2$ & $\frac{J}{2}$ & $\frac{J+2}{2}$ & $\rep{1}$ \\[4pt]
    $(\phi^{++})^J \chi_{\sR}^{+a} \chi_{\sL}^{+b}$ & $J+1$ & $\frac{J+1}{2}$ & $\frac{J+1}{2}$ & $\rep{1}+\rep{3}$ \\[4pt]
    $(\phi^{++})^J \chi_{\sL}^{++} \chi_{\sL}^{+-}$ & $J$ & $\frac{J+2}{2}$ & $\frac{J}{2}$ & $\rep{1}$ \\
    \midrule
    $(\phi^{++})^J \chi_{\sR}^{++} \chi_{\sR}^{+-} \chi_{\sL}^{+\pm}$ & $J+2$ & $\frac{J+1}{2}$ & $\frac{J+2}{2}$ & $\rep{2}$ \\[4pt]    
    $(\phi^{++})^J \chi_{\sR}^{+\pm} \chi_{\sL}^{++} \chi_{\sL}^{+-}$ & $J+1$ & $\frac{J+2}{2}$ & $\frac{J+1}{2}$ & $\rep{2}$ \\
    \midrule
    $(\phi^{++})^J \chi_{\sR}^{++} \chi_{\sR}^{+-} \chi_{\sL}^{++} \chi_{\sL}^{+-}$ & $J+2$ & $\frac{J+2}{2}$ & $\frac{J+2}{2}$ & $\rep{1}$ \\
    \bottomrule
  \end{tabular}

  \caption{The sixteen 1/2-BPS states containing $J$ copies of the field $\phi^{++}$ obtained by inserting the four fermionic zero modes. The length $L$ of the corresponding spin-chain ground state and the charges under $\algSU(2)_{\sL}$, $\algSU(2)_{\sR}$ and the $\algSU(2)_{\circ}$ representation are also given. \label{tab:spin-chain-BPS-states-J}}
\end{table}

It is now straightforward to count the total number of states of a given charge $(J_{\sL}, J_{\sR})$ that appear. For $J \geq 2$ we find
\begin{itemize}
\item 6 bosonic states with charges $(\tfrac{J}{2}, \tfrac{J}{2})$,
\item 4+4 fermionic states with charges $(\tfrac{J-1}{2}, \tfrac{J}{2})$ and $(\tfrac{J}{2}, \tfrac{J-1}{2})$,
\item 1+1 bosonic states with charges $(\tfrac{J}{2}-1, \tfrac{J}{2})$ and $(\tfrac{J}{2}, \tfrac{J}{2}-1)$.
\end{itemize}
Additionally, there are
\begin{itemize}
\item 2+2 fermionic states with charges $(\tfrac{1}{2}, 0)$ and $(0, \tfrac{1}{2})$
\item 5 bosonic states with charges $(\tfrac{1}{2}, \tfrac{1}{2})$ .
\end{itemize}
Summing over all the single-particle and multi-particle states with the terms weighted as in the Poincar\'e polynomial in equation~\eqref{eq:poincare-polynomial}, we find that the partition function in the 1/2-BPS sector of the spin chain perfectly reproduces the result for the symmetric orbifold $\SymN(\Torus^4)$ presented in equation \eqref{eq:ppt4}.

As anticipated in section \ref{sec:CFT}, the computation of the (modified) elliptic genus requires more care, since it diverges in the large-$N$ limit. Following \cite{deBoer:1998us,Maldacena:1999bp}, we assign to each state in table \ref{tab:spin-chain-BPS-states-J} a degree $d = J+1$. One can then compute the elliptic genus $\tilde{\mathcal{E}}_2(Q,q,y)$ as
\begin{equation}
  \label{eq:megspinchain}
  \tilde{\mathcal{E}}_2(Q,q,y) = \mathrm{Tr}_{|\textrm{anything}\rangle\sL \otimes |\textrm{chiral primary}\rangle\sR} (-1)^F Q^d q^{2 \gen{D}\sL} y^{2 \gen{J}\sL} (2 \gen{J}\sR)^2 .
\end{equation}
In appendix~\ref{app:elliptic-genus} we show that if we consider states with (left) conformal dimension $D\sL < d/4$, the result is given by
\begin{equation}
  \tilde{\mathcal{E}}_2(Q,q,y) = \frac{2Q}{(1-Q)^2} + \ldots = \sum_N 2N Q^N + \ldots ,
\end{equation}
in perfect agreement with \eqref{eq:leadingeg}.

\section{Protected states for  \texorpdfstring{$\AdS_3 \times \Sphere^3 \times \Sphere^3 \times \Sphere^1$}{AdS3 x S3 x S3 x S1}}
\label{sec:s1}

Another interesting string theory background is  $\AdS_3 \times \Sphere^3 \times \Sphere^3 \times \Sphere^1$, which also preserves sixteen supercharges. Strings in this background are expected to be dual to a CFT with \emph{large} $\superN=(4,4)$ superconformal symmetry. 
As discussed in section~\ref{sec:CFT}, the global part of this symmetry is given by the algebra $\algD{\alpha}_{\sL} \oplus \algD{\alpha}_{\sR}$, where the parameter $\alpha$ is related to the ratio of the radii of the two three-spheres
\begin{equation}
\frac{\alpha}{1-\alpha}=\frac{R^2_{\Sphere^3_-}}{R^2_{\Sphere^3_+}}\,.
\end{equation}

The all-loop integrable S matrix for $\AdS_3 \times \Sphere^3 \times \Sphere^3 \times \Sphere^1$ was found in~\cite{Borsato:2015mma}. The world-sheet spectrum contains, among other excitations, two massless fermions. It is straightforward to generalize the all-loop derivation of the protected spectrum carried out in section~\ref{sec:fermioniczeromodes} to the present background. As before, the Bethe equations have a (trivial) BMN vacuum solution 
above which there are excitations that are obtained by acting with Zamolodchikov-Fadeev creation operators in the conventional way. In general, their momenta are determined via non-trivial solutions to the Bethe equations and so will have non-zero energies. However, just as in section~\ref{sec:fermioniczeromodes}, \emph{zero-momentum} solutions of the all-loop Bethe equations corresponding to massless fermionic excitations above the BMN vacuum can be obtained very straightforwardly, because the Bethe equations reduce to $1=1$ at zero momentum. This demonstrates that to all orders in the string tension, the massless fermionic excitations are gapless just as in the case of the $\AdS_3 \times \Sphere^3 \times\Torus^4$ background. 

Let us discuss the protected $\AdS_3 \times \Sphere^3 \times \Sphere^3 \times \Sphere^1$ spectrum in more detail from a spin-chain perspective. 
The spin chain describing the massive sector of the pure-RR $\AdS_3 \times \Sphere^3 \times \Sphere^3 \times \Sphere^1$ integrable system was constructed in~\cite{OhlssonSax:2011ms}. 
The basic building blocks of the spin chain are the two \emph{doublet} representations of $\algD{\alpha}$. These representations carry $\algSL(2) \oplus \algSU(2) \oplus \algSU(2)$ charges $(D,J_+,J_-)=(\tfrac{\alpha}{2} ,\tfrac{1}{2},0)$ and $(\tfrac{1-\alpha}{2},0,\tfrac{1}{2})$, and are 1/4-BPS since they saturate the bound discussed in section~\ref{sec:CFT}
\begin{equation}\label{eq:d21a-BPS-bound}
  D \geq \alpha J_+ + (1-\alpha) J_- .
\end{equation}
To write down the supersymmetric BMN ground states we introduce two fields $\Phi_{\alpha}$ and $\Phi_{1-\alpha}$ which are the highest weight states of the representations $(\tfrac{\alpha}{2} ,\tfrac{1}{2},0) \otimes (\tfrac{\alpha}{2} ,\tfrac{1}{2},0)$ and $(\tfrac{1-\alpha}{2},0,\tfrac{1}{2}) \otimes (\tfrac{1-\alpha}{2},0,\tfrac{1}{2})$ of $\algD{\alpha}_{\sL} \oplus \algD{\alpha}_{\sR}$. The supersymmetric BMN ground states then take the form
\begin{equation}
  \ket{ ( \Phi_{\alpha} \Phi_{1-\alpha} )^L } .
\end{equation}
Note that the massive spin chain is alternating, with even and odd sites transforming in different representations of the symmetry algebra.
Each doublet representation preserves half of the 16 supersymmetries, but the tensor product of the two representations only preserves four supersymmetries so that the spin-chain ground states are~1/4-BPS.
This is to be expected: while string theory on  $\AdS_3 \times \Sphere^3 \times \Sphere^3 \times \Sphere^1$ preserves 16 supersymmetries like $\AdS_3 \times \Sphere^3 \times \Torus^4$, fixing light-cone gauge here breaks twelve supersymmetries, rather than the eight that are broken in $\AdS_3 \times \Sphere^3 \times \Torus^4$~\cite{Babichenko:2009dk}.
  
The  two massless fermionic excitations correspond to sites transforming in representations of the form $\rep{1} \otimes (\tfrac{1}{2},\tfrac{1}{2},\tfrac{1}{2})$  and $(\tfrac{1}{2},\tfrac{1}{2},\tfrac{1}{2}) \otimes \rep{1}$.\footnote{%
  The $\algD{\alpha}$ representation $(\tfrac{1}{2},\tfrac{1}{2},\tfrac{1}{2})$ is the only 1/4-BPS representation that appears in the decomposition of the tensor product of the two doublet representations $(\tfrac{\alpha}{2} ,\tfrac{1}{2},0)$ and $(\tfrac{1-\alpha}{2},0,\tfrac{1}{2})$.%
} %
Denoting the highest weight state of these representations by $\chi_{\sL}$ and $\chi_{\sR}$, the chiral ring is then made out of states of the form
\begin{equation}
  \ket{ ( \Phi_{\alpha} \Phi_{1-\alpha} )^L } , \quad
  \ket{ ( \Phi_{\alpha} \Phi_{1-\alpha} )^L \chi_{\sL} } , \quad
  \ket{ ( \Phi_{\alpha} \Phi_{1-\alpha} )^L \chi_{\sR} } , \quad
  \ket{ ( \Phi_{\alpha} \Phi_{1-\alpha} )^L \chi_{\sL} \chi_{\sR} }.
\end{equation}
These states saturate~\eqref{eq:d21a-BPS-bound} and have equal angular momentum on the two three-spheres, $J_+ = J_-$. Moreover, the four $\algD{\alpha}_{\sL} \oplus \algD{\alpha}_{\sR}$ multiplets precisely fit into one multiplet of the large $\mathcal{N}=(4,4)$ algebra.

This prediction, valid for generic string tension, is at odds with the proposal for the supergravity spectrum of ref.~\cite{deBoer:1999gea}. There it was proposed that protected states with $J_+\neq J_-$ should exists, yielding a much larger degeneracy; this was derived under the assumption that all $\AdS_3 \times \Sphere^3 \times \Sphere^3 \times \Sphere^1$ Ka\l{}u\.za-Klein modes sit in short representations of 
$\algD{\alpha}$---an assumption that may indeed lead to overestimating the number of protected states.\footnote{We would like to thank Jan de Boer for detailed discussions related to this point.}
This additional degeneracy appeared unlikely to exist at generic points of the moduli space, as emphasised in ref.~\cite{Gukov:2004ym}. In fact, states with $J_+\neq J_-$ must satisfy different shortening conditions in 
$\algD{\alpha}$ and in the large $\superN=(4,4)$ superconformal algebra, which made it difficult to believe that they would remain protected. Our derivation of the protected spectrum explicitly confirms the reasoning of~\cite{Gukov:2004ym}. It is however still interesting to see whether such an additional degeneracy is there at all at the supergravity point. 

A simple yet effective way to investigate the protected supergravity spectrum is to consider point-like string solutions in $\AdS_3 \times \Sphere^3 \times \Sphere^3 \times \Sphere^1$. These are in one-to-one correspondence with bosonic supersymmetric ground states: should these exist at $J_+\neq J_-$, so must such classical solutions. Point-like string solutions for $\AdS_3 \times \Sphere^3 \times \Sphere^3 \times \Sphere^1$ were briefly discussed in~\cite{Babichenko:2009dk} (see also~\cite{Lloyd:2013wza,Abbott:2014rca}), and  we review them here.
The strings can be taken to move only along the time direction $t$ of $\AdS_3$ and the great circles $\varphi_{\pm}$ of the two 
three-spheres. The equations of motion are then solved by
\begin{equation}
  t = \kappa\tau , \quad
  \varphi_+ = \omega_+\tau , \quad
  \varphi_- = \omega_-\tau ,
\end{equation}
where the constants $\kappa$ and $\omega_{\pm}$ are related to the energy $D$ and angular momenta $J_{\pm}$ by
\begin{equation}
  D = \kappa , \quad
  J_+ = \alpha \omega_+ , \quad
  J_- = (1-\alpha) \omega_- .
\end{equation}
The Virasoro constraint gives
\begin{equation}\label{eq:point-like-energy}
  D = \sqrt{\alpha J_+^2 + (1-\alpha) J_-^2},
\end{equation}
and as discussed the 1/4-BPS bound of $\algD{\alpha}_{\sL} \oplus \algD{\alpha}_{\sR}$ takes the form~\eqref{eq:d21a-BPS-bound}.
We see immediately that the energies~\eqref{eq:point-like-energy} only saturate the bound when $J_+ = J_-$. 
The states that saturate the bound, together with the two fermionic zero modes that we obtained above, 
then generate the spectrum of protected operators,\footnote{%
  Here we denote a short multiplet of $\algD{\alpha}_{\sL} \oplus \algD{\alpha}_{\sR}$ by the charges of its highest weight state under the compact subalgebra as $[ J_+^{\sL} , J_-^{\sL} ; J_+^{\sR} , J_-^{\sR}  ]_s$.
}%
\begin{equation}\label{eq:s1-chiral-ring}
  \begin{aligned}
    \bigoplus_J \bigl(&
    [ J , J ; J , J ]_s \oplus [ J + \tfrac{1}{2}, J + \tfrac{1}{2} ; J + \tfrac{1}{2}, J + \tfrac{1}{2} ]_s
    \\ &\oplus [ J , J ; J + \tfrac{1}{2}, J + \tfrac{1}{2} ]_s \oplus [ J + \tfrac{1}{2}, J + \tfrac{1}{2} ; J , J ]_s \bigr) .
  \end{aligned}
\end{equation}
For $J_+ \neq J_-$ the point-like strings do not saturate the BPS bound and hence fall in long representations. These solutions correspond to supergravity modes with classical energy~\eqref{eq:point-like-energy}. 
In addition, since they sit in long representations, their energy will receive corrections in the string tension.

\section{Protected states in mixed-flux \texorpdfstring{$\AdS_3$}{AdS3} backgrounds}
\label{sec:mixed}

So far we have focused on string theory on $\AdS_3$ backgrounds supported by a RR three-form. Both $\AdS_3\times\Sphere^3\times\Sphere^3\times\Sphere^1$
 and $\AdS_3\times\Sphere^3\times\Torus^4$  remain classsically integrable when additionally the NSNS three-form is turned on~\cite{Cagnazzo:2012se}. Moreover, up to the dressing factors, the integrable S~matrices can also be fixed entirely by symmetry in these cases~\cite{Lloyd:2014bsa,Borsato:2015mma}. In fact, for both families of backgrounds the structure of the worldsheet fluctuations and the symmetries of the light-cone gauge-fixed theory remain the same when the NSNS flux is turned on. The only modification is that the representation of the light-cone gauge symmetries gets deformed in a way that can be accounted for by a suitable deformation of the Zhukovski parameters.\footnote{%
The crossing equations are also modified in the presence of NSNS flux.%
}
This means that the form of the Bethe equations is independent of the flux, which only enters in the relation between the Bethe roots and the world-sheet energy and momentum. The dispersion relation takes the form~\cite{Hoare:2013lja,Lloyd:2014bsa}
\begin{equation}
\label{eq:mixeddispersion}
E(p)=\sqrt{(m-\tfrac{k}{2\pi} p)^2+4h^2\sin^2\tfrac{p}{2}}
\end{equation}
where $k$ is the integer-valued coupling of the Wess-Zumino term of the bosonic string NLSM action, and the coupling $h$ is suitably rescaled~\cite{Lloyd:2014bsa,Borsato:2015mma}.

Without needing to resort to constructing the spin chain, we immediately see that when $m=0$ we will have again gapless modes in the spectrum; in particular, just like in the pure-RR case, we find four fermionic zero modes over the BMN vacuum for the  $\AdS_3\times\Sphere^3\times\Torus^4$ background and two for $\AdS_3\times\Sphere^3\times\Sphere^3\times\Sphere^1$. Their quantisation leads to a tower of sixteen 1/2-BPS states for each ground state in the former case, and of four 1/4-BPS states in the latter, so that the structure of the chiral ring is the same for any flux.

It is worth remarking that, even if the structure of the Bethe equations is not modified by $k \geq 1$, it may very well be that the spin-chain  description is altered rather drastically. There are two reasons for this. Firstly, the mixed-flux dispersion relation~\eqref{eq:mixeddispersion} is no longer a periodic function of the world-sheet momentum~$p$, which might suggest that the discretised lattice picture should be altered. Secondly, we expect the spin-chain picture to appear when the coupling constant is small. However, in the mixed flux case, the string tension is bounded from below due to the presence of the Wess-Zumino term in the NLSM action, and hence there is no genuine weak-coupling limit in the dual CFT. It would be very interesting to investigate this further.

\section{Wrapping corrections}
\label{sec:wrapping}

It is well-known~\cite{Ambjorn:2005wa} that the Bethe equations for AdS/CFT integrability yield an incomplete description of the spectrum, as they do not account for finite-size effects of the type first described by L\"uscher for relativistic systems~\cite{Luscher:1985dn,Luscher:1986pf}. These corrections can be interpreted as arising from particles wrapping the worldsheet cylinder one or more times, and can be in principle be understood as part of a systematic expansion around large volume~\cite{Bajnok:2008qj,Bombardelli:2013yka}. The finite-volume description that resums  all these effects can be expressed by employing more refined formalisms such as the mirror thermodynamic Bethe ansatz (TBA)~\cite{Arutyunov:2007tc, Arutyunov:2009zu, Gromov:2009tv, Bombardelli:2009ns, Arutyunov:2009ur} or the quantum 
spectral curve~\cite{Gromov:2013pga,Gromov:2014caa,Bombardelli:2017vhk}.
It is interesting to note that generically such wrapping corrections may lift degenerate multiplets~\cite{Sfondrini:2011rr}. Therefore, we should ask ourselves whether the spectrum of protected states we found from the Bethe equations is robust when wrapping effects are taken into account.%
\footnote{We thank the anonymous referee for raising this point, and giving us the opportunity to report on the results presently.}

As a mirror-TBA formalism for AdS/CFT pairs involving gapless excitation has not yet been developed, discussing the exact finite-volume spectrum is beyond our reach. Interestingly, we will nonetheless be able to see exactly what happens to protected states. Let us start by recalling the form of a single-wrapping correction%
\footnote{See also \cite{Janik:2010kd,Bombardelli:2016rwb}}
\begin{equation}
\delta D_{\text{wrapping}}\approx
\int \text{d}\tilde{p}\ e^{-L \tilde{E}} 
\sum_{a} (-1)^{F_a} S^{ab_1}(\tilde{p},p_1)\cdots S^{ab_k}(\tilde{p},p_k)\,.
\end{equation}
Here we consider a physical state consisting of $k$ particles with fixed momenta $p_1,\dots p_k$ and flavours $b_1,\dots b_k$. Moreover, we consider the scattering of a virtual particle in the the so-called mirror kinematics~\cite{Arutyunov:2007tc} with mirror momentum $\tilde{p}$, mirror energy $\tilde{E} $, and flavour $a$. We will have to sum over all possible particle types $a$ and integrate over momenta $\tilde{p}$. Note that we have assumed that the scattering is always given by a pure transmission process, for reasons that will become clear soon. Let us now take the the case in which the index $a$ labels the states in a given representation~$\rho$; we recognise the mirror transfer matrix
\begin{equation}
T_{\rho}(\tilde{p};\{p_1,\dots p_k\})=\sum_{a\in\rho} (-1)^{F_a} S^{ab_1}(\tilde{p},p_1)\cdots S^{ab_k}(\tilde{p},p_k).
\end{equation}
We will then have to sum over all possible representations $\rho$, for all fundamental particles as well as bound states. Let us now specialise the case in which all particles of flavours $b_1,\dots b_k$ are massless modes at zero momentum $p_1=\dots =p_k=0$. Recall that such particles are \emph{singlets} of the light-cone symmetry algebra $\alg{psu}(1|1)^4_{\text{c.e.}}$~\cite{Borsato:2014hja}. Hence the scattering that we are considering is
\begin{equation}
S:\quad \rho \otimes 1 \mapsto	 1 \otimes \rho\,,
\end{equation}
\textit{i.e.}\ it can only result in pure transmission. Furthermore, invariance under $\alg{psu}(1|1)^4_{\text{c.e.}}$ means that all states in the multiplet $\rho$ will pick up the same phase shift, $S^{ab}(\tilde{p},0)= \exp[{i \varphi^{b}(\tilde{p})}]$, independently from $a$.%
\footnote{When $\rho$ is a single-particle representation, this can be readily verified from the explicit formulae given in refs.~\cite{Borsato:2014hja,Borsato:2015mma}; S-matrix fusion then yields all scattering phases.}
All in all we find
\begin{equation}
T_{\rho}(\tilde{p};\{p_1,\dots p_k\})=\exp\big[{i \varphi^{b_1}(\tilde{p})+ \cdots i \varphi^{b_k}(\tilde{p})}\big]\ \sum_{a\in\rho} (-1)^{F_a}=0,
\end{equation}
where we used the fact that all representations~$\rho$ are supersymmetric. This cancellation process is completely analogous to the one that guarantees that the BMN vacuum is protected from wrapping corrections. In fact, while it is subtle to systematically define multiple-wrapping formulae, \textit{cf.}\ also~\cite{Heller:2008at,Bombardelli:2013yka,Abbott:2015pps}, it is seems natural that the wrapping corrections for the states that we consider here should follow the same fate as those for the BMN vacuum---this can be explicitly checked for certain wrapping contributions for which explicit formulae are available~\cite{Bombardelli:2013yka,Abbott:2015pps}.

As we have seen, supersymmetry forces the wrapping corrections to vanish order by order for the protected states that we constructed out of fermionic zero-modes. In fact, our argument applies to both massless \emph{fermionic and bosonic} zero-modes. In the latter case, it shows that the shift isometries along the flat directions are unbroken as expected.
 The argument holds for both the  $\AdS_3\times\Sphere^3\times\Torus^4$ and the $\AdS_3\times\Sphere^3\times\Sphere^3\times \Sphere^1$ backgrounds, in presence of arbitrary three-form fluxes. Therefore, this condition may be taken a useful guiding principle in the formulation of a set of mirror TBA equations for the finite-volume spectrum: each protected state should yield a sector for the mirror TBA equations. The exact Bethe roots will be fixed in a manner similar to what happens in ref.~\cite{Arutyunov:2012tx} for ``exceptional'' operators; furthermore, and in contrast to the case of exceptional operators, the vacuum energy should be quantised in each sector.

\section{Conclusion}
\label{sec:conclusions}

In this paper we investigated the spectrum of closed-string excitations in integrable $\AdS_3/\CFT_2$ backgrounds.  We developed a spin chain for the weak-coupling limit of the all-loop Bethe equations~\cite{Borsato:2016kbm} of string theory on $\AdS_3\times\Sphere^3\times\Torus^4$ supported by RR fluxes, emphasizing the role that length-changing interactions play in it. The spin-chain constructed here is very similar to the one appearing in large-$N$ perturbative calculations in the Higgs-branch CFT~\cite{Sax:2014mea}. It would be interesting to extend the perturbative analysis carried out there to the complete weak-coupling theory and match with the spin~chain proposed here. Further, constructing a framework for incorporating wrapping interaction~\cite{Ambjorn:2005wa,Abbott:2015pps} into $\AdS_3$ holography, such as the mirror TBA~\cite{Arutyunov:2007tc, Arutyunov:2009zu, Gromov:2009tv, Bombardelli:2009ns, Arutyunov:2009ur} and quantum 
spectral curve~\cite{Gromov:2013pga,Gromov:2014caa,Bombardelli:2017vhk} remains an important task.

We showed that protected closed-string states are determined by classifying \emph{zero-momentum} Bethe roots. This can be done straightforwardly at all values of the string tension in the planar theory. We found that, in addition to the BMN vacuum, a number of further protected states exist. For example, in the weakly-coupled spin-chain description 
of the pure-RR $\AdS_3\times\Sphere^3\times\Torus^4$ theory, these are obtained by inserting up to four fermionic zero modes, yielding a $16$-fold degeneracy for each ferromagnetic vacuum.%
\footnote{%
Our results represent a derivation of the proposal for incorporating massless modes into the integrable spin-chain given in~\cite{Sax:2012jv} which was based on the $\alpha\rightarrow 0$ limit of massive modes in the $\AdS_3\times\Sphere^3\times\Sphere^3\times \Sphere^1$ theory.%
}
We have shown that our all-loop integrability-based analysis of protected states matches precisely the results of~\cite{deBoer:1998kjm} where the supergravity spectrum was found to agree with the large $N$ limit of the $\SymN(\Torus^4)$ orbifold, and with the modified elliptic genus~\cite{Maldacena:1999bp}. This indicates that the weakly-coupled spin chain lives in the same moduli~space as the symmetric-product orbifold CFT. It would be important to pinpoint where exactly the weakly-coupled spin chain point is, and its relation with the symmetric-product orbifold.
It is intriguing to note that the generators of the chiral ring for the weakly-coupled spin chain collected in table~\ref{tab:spin-chain-BPS-states-J} closely resemble their counterparts in the symmetric-product orbifold CFT. While this guarantees that the spectrum of 1/2-BPS operators is identical, it is hard to say whether we should expect a one-to-one matching of the generators. Lessons from supergravity~\cite{Taylor:2007hs} suggest that we should expect a mixing between single- and multi-particle (or, in our language, single- and multi-trace) states. A way to unravel this question would be to match the three-point functions of protected operators~\cite{Lunin:2000yv,Lunin:2001pw,Gaberdiel:2007vu, Dabholkar:2007ey, Pakman:2007hn, Taylor:2007hs}, which are also constant on the moduli space~\cite{deBoer:2008ss,Baggio:2012rr}.

It is also interesting to notice that this analysis, valid for generic values of the tension, would be significantly different in the strict $h\to0$ limit. Here the anomalus dimension $\delta D$ would vanish identically and we would find a plethora of accidental 1/2- and 1/4-BPS states. While it is hardly surprising that a ``free'' point of the moduli space might exhibit accidental symmetries (as it is also the case in $\mathcal{N}=4$ SYM, \cf~\eg~\cite{Beisert:2004di,Beisert:2004ry}),  this is a qualitative difference with the symmetric-product orbifold CFT, where no such extra degeneracy exists. Furthermore, from a sigma-model perspective, BPS states correspond to cohomology classes of the target manifold; this suggests that the weakly-coupled spin~chain corresponds to some singular limit of the target space geometry. It would be interesting to explore this in more detail.

We then turned to the $\AdS_3\times\Sphere^3\times\Sphere^3\times \Sphere^1$ background. An analysis of the zero-momentum Bethe roots shows that the protected closed-string spectrum of this theory is constructed out of BMN vacua and two gapless fermionic excitations. The results are again valid for any value of the string tension in the planar theory. The fermionic zero modes give rise to a $4$-fold degeneracy for each ferromagnetic BMN vacuum which precisely yields a multiplet of the large $\mathcal{N}=(4,4)$ algebra.  In particular, as we have confirmed by studying classical point-like string solutions, only states with equal $\alg{su}(2)$ charge with respect to the two three-spheres are protected. This is a significantly smaller degeneracy than the one proposed in supergravity in~\cite{deBoer:1999gea}, which was derived under the assumption that all KK modes sit in short multiplets. 
The potential for a smaller protected spectrum than the one proposed in~\cite{deBoer:1999gea} was already argued for in~\cite{Gukov:2004ym}
based on the different shortening conditions in the super-Lie and super-Virasoro algebras. It would be interesting to determine whether this new understanding of the protected spectrum can shed any light on the long-standing problem of identifying the dual of $\AdS_3\times\Sphere^3\times\Sphere^3\times \Sphere^1$. In this regard, it is worth noting that the spectrum that we found is compatible with that of symmetric-product orbifold CFTs with large $\mathcal{N}=(4,4)$ symmetry, such as the ones of class $\mathcal{S}_{\kappa}$ considered in reference~\cite{Gukov:2004ym}.

Finally, the $\AdS_3$ backgrounds considered here can be supported by a mixture of RR and NSNS flux all of which are known to be integrable. Based on the integrability results we have argued that the spectrum of protected operators remains the same as one deforms away from the pure-RR case. This is quite interesting, as exchanging these background fluxes amounts to an S-duality transformation, which in general would affect rather drastically the spectrum. The pure NSNS backgrounds can be understood using WZW techniques~\cite{Elitzur:1998mm,Maldacena:2000hw} and it would be interesting to see what the relation between these and the integrable methods used here is.

\section*{Acknowledgements}

We would like to thank Lorenz Eberhardt, Matthias Gaberdiel, Rajesh Gopakumar and Wei Li for sharing with us their manuscript~\cite{Eberhardt:2017fsi} before it appeared on the arXiv, and for their comments on our results.
We would like to thank Riccardo Borsato for interesting discussions on many aspects of integrability in $\AdS_3/\CFT_2$, Jan de Boer for explanations and discussions of the supergravity analysis of protected spectra in $\AdS_3$ backgrounds, Andrea Prinsloo for discussions of giant gravitons in $\AdS_3$ backgrounds, and Kostya Zarembo for numerous discussions.
Finally, we would like to thank the Galileo Galilei Institute in Florence for their kind hospitality during part of this work.
A.S.\@ would like to thank Andrea Dei, Lorenz Eberhardt, Matthias Gaberdiel, Edi Gava, Kumar Narain, and Ida Zadeh for interesting discussions on $\AdS_3/\CFT_2$.
B.S.\@ would like to thank Kelly Stelle for enlightening discussions on KK supergravity modes.
The work of M.B.\@ was supported in part by the European Research Council grant no.~ERC-2013-CoG 616732 HoloQosmos and in part by the FWO and the European Union's Horizon 2020 research and innovation programme under the Marie Sk{\l}odowska-Curie grant agreement no.~665501. M.B.\@ is an FWO [PEGASUS]${}^2$ Marie Sk{\l}odowska-Curie Fellow.
O.O.S.\@ was supported by ERC Advanced grant No.\@ 341222.
A.S.'s research was partially supported by the NCCR SwissMAP, funded by the Swiss National Science Foundation.
B.S.\@ acknowledges funding support from an STFC Consolidated Grant ``Theoretical Physics at City University'' ST/J00037X/1.
A.T.\@ thanks the EPSRC for funding under the First Grant project EP/K014412/1 and the STFC under the Consolidated Grant project nr.~ST/L000490/1.

\appendix

\section{The \texorpdfstring{$\algPSU(1,1|2)$}{psu(1,1|2)} superalgebra}
\label{sec:psu112}

The $\algSU(1,1|2)$ superalgebra consists of the $\algSU(2)$ R-symmetry generators $\gen{J}^{\alpha}{}_{\beta}$, the translation $\gen{P}$, the supercharges $\gen{Q}_{\alpha}$ and $\dot{\gen{Q}}^{\alpha}$, the dilatation $\gen{D}$, the special conformal transformation $\gen{K}$, the conformal supercharges $\gen{S}^{\alpha}$ and $\dot{\gen{S}}_{\alpha}$ and the central charge $\gen{C}$. We write the $\algSU(2)$ commutation relations as\footnote{%
  Up to a change in the convention for the indices this follows the notation of appendix~D of~\cite{Beisert:2004ry}.%
}%
\begin{equation}
  \comm{\gen{J}^{\alpha}{}_{\beta}}{\gen{R}^{\gamma}} = + \delta^{\gamma}_{\beta} \gen{R}^{\alpha} - \tfrac{1}{2} \delta^{\alpha}_{\beta} \gen{R}^{\gamma} , \qquad
  \comm{\gen{J}^{\alpha}{}_{\beta}}{\gen{R}_{\gamma}} = - \delta^{\alpha}_{\gamma} \gen{R}_{\beta} + \tfrac{1}{2} \delta^{\alpha}_{\beta} \gen{R}_{\gamma} , 
\end{equation}
where $\gen{R}$ is an arbitrary generator. The $\algSU(1,1)$ algebra takes the form
\begin{equation}
  \comm{\gen{K}}{\gen{P}} = 2\gen{D} , \qquad
  \comm{\gen{D}}{\gen{P}} = + \gen{P} , \qquad
  \comm{\gen{D}}{\gen{K}} = - \gen{K} .
\end{equation}
The action of $\gen{P}$ and $\gen{K}$ on the supercharges is given by
\begin{equation}
  \comm{\gen{K}}{\gen{Q}_{\alpha}} = + \dot{\gen{S}}_{\alpha} , \qquad
  \comm{\gen{K}}{\dot{\gen{Q}}^{\alpha}} = + \gen{S}^{\alpha} , \qquad
  \comm{\gen{P}}{\gen{S}^{\alpha}} = - \dot{\gen{Q}}^{\alpha} , \qquad
  \comm{\gen{P}}{\dot{\gen{S}}_{\alpha}} = - \gen{Q}^{\alpha} ,
\end{equation}
and the non-trivial anti-commutators by
\begin{equation}
  \begin{aligned}
    \acomm{\gen{Q}_{\alpha}}{\dot{\gen{Q}}^{\beta}} &= \delta_{\alpha}^{\beta} \gen{P} , \qquad &
    \acomm{\gen{Q}_{\alpha}}{\gen{S}^{\beta}} &= + \gen{J}^{\beta}{}_{\alpha} + \delta^{\beta}_{\alpha} ( \gen{D} - \gen{C} ) , \\
    \acomm{\gen{S}^{\alpha}}{\dot{\gen{S}}_{\beta}} &= \delta^{\alpha}_{\beta} \gen{K} , \qquad &
    \acomm{\dot{\gen{Q}}^{\alpha}}{\dot{\gen{S}}_{\beta}} &= - \gen{J}^{\alpha}{}_{\beta} + \delta^{\alpha}_{\beta} ( \gen{D} + \gen{C} ) .
  \end{aligned}
\end{equation}
The (conformal) supercharges carry dimension ($\gen{D}$) 
\begin{equation}
  \operatorname{dim}(\gen{Q}) = +\tfrac{1}{2} , \qquad
  \operatorname{dim}(\dot{\gen{Q}}) = +\tfrac{1}{2} , \qquad
  \operatorname{dim}(\gen{S}) = -\tfrac{1}{2} , \qquad
  \operatorname{dim}(\dot{\gen{S}}) = -\tfrac{1}{2} .
\end{equation}
If we restrict ourselves to representations with a vanishing central charge $\gen{C}$ the resulting algebra is $\algPSU(1,1|2)$.

\paragraph{Automorphism.}

The algebra $\algPSU(1,1|2)$ has an automorphism which we will refer to as $\algSU(2)_{\bullet}$ with generators $\gen{J}_\bullet$ and $\gen{J}_\bullet^{\pm}$ which satisfy
\begin{equation}
  \begin{gathered}
    \begin{aligned}
    \comm{\gen{J}_{\bullet}}{\gen{J}_{\bullet}^{\pm}} &= \pm \gen{J}_{\bullet}^{\pm} , \qquad &
    \comm{\gen{J}_{\bullet}^+}{\gen{J}_{\bullet}^-} &= 2 \gen{J}_{\bullet} , \\
      \comm{\gen{J}_{\bullet}^+}{\dot{\gen{Q}}^{\alpha}} &= \epsilon^{\alpha\beta} \gen{Q}_{\beta} , \qquad &
      \comm{\gen{J}_{\bullet}^+}{\gen{S}^{\alpha}} &= \epsilon^{\alpha\beta} \dot{\gen{S}}_{\beta} , \\
      \comm{\gen{J}_{\bullet}^-}{\gen{Q}_{\alpha}} &= \epsilon_{\alpha\beta} \dot{\gen{Q}}^{\beta} , \qquad &
      \comm{\gen{J}_{\bullet}^-}{\dot{\gen{S}}_{\alpha}} &= \epsilon_{\alpha\beta} \gen{S}^{\beta} ,
    \end{aligned} \\
    \comm{\gen{J}_{\bullet}}{\gen{Q}_{\alpha}} = +\tfrac{1}{2} \gen{Q}_{\alpha} , \quad
    \comm{\gen{J}_{\bullet}}{\dot{\gen{Q}}^{\alpha}} = -\tfrac{1}{2} \dot{\gen{Q}}^{\alpha} , \quad
    \comm{\gen{J}_{\bullet}}{\dot{\gen{S}}^{\alpha}} = +\tfrac{1}{2} \dot{\gen{S}}^{\alpha} , \quad
    \comm{\gen{J}_{\bullet}}{\gen{S}_{\alpha}} = -\tfrac{1}{2} \gen{S}_{\alpha} .
  \end{gathered}
\end{equation}

\paragraph{Grading.}

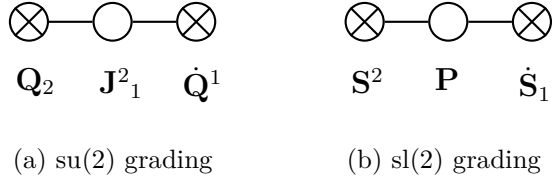
\begin{figure}
  \centering

  \subfloat[$\algSU(2)$ grading\label{fig:dynkin-su22-su}]{
    \begin{tikzpicture}
      [
      thick,
      node/.style={shape=circle,draw,thick,inner sep=0pt,minimum size=5mm}
      ]

      \useasboundingbox (-1.5cm,-1.5cm) rectangle (1.5cm,1cm);

      \node (v1) at (-1.1cm, 0cm) [node] {};
      \node (v2) at (  0.0cm, 0cm) [node] {};
      \node (v3) at (  1.1cm, 0cm) [node] {};

      \draw (v1.south west) -- (v1.north east);
      \draw (v1.north west) -- (v1.south east);

      \draw (v3.south west) -- (v3.north east);
      \draw (v3.north west) -- (v3.south east);

      \draw (v1) -- (v2);
      \draw (v2) -- (v3);

      \node at ($(v1.south)+(0,-0.2cm)$) [anchor=north] {$\mathrlap{\gen{Q}_2}\phantom{\gen{Q}}$};
      \node at ($(v2.south)+(0,-0.2cm)$) [anchor=north] {$\mathrlap{\gen{J}^2{}_1}\phantom{\gen{Q}}$};
      \node at ($(v3.south)+(0,-0.2cm)$) [anchor=north] {$\mathrlap{\dot{\gen{Q}}^1}\phantom{\gen{Q}}$};
    \end{tikzpicture}
  }
  \hspace{1cm}
  \subfloat[$\algSL(2)$ grading \label{fig:dynkin-su22-sl}]{
    \begin{tikzpicture}
      [
      thick,
      node/.style={shape=circle,draw,thick,inner sep=0pt,minimum size=5mm}
      ]

      \useasboundingbox (-1.5cm,-1.5cm) rectangle (1.5cm,1cm);

      \node (v1) at (-1.1cm, 0cm) [node] {};
      \node (v2) at (  0.0cm, 0cm) [node] {};
      \node (v3) at (  1.1cm, 0cm) [node] {};

      \draw (v1.south west) -- (v1.north east);
      \draw (v1.north west) -- (v1.south east);

      \draw (v3.south west) -- (v3.north east);
      \draw (v3.north west) -- (v3.south east);

      \draw (v1) -- (v2);
      \draw (v2) -- (v3);

      \node at ($(v1.south)+(0,-0.2cm)$) [anchor=north] {$\mathrlap{\gen{S}^2}\phantom{\gen{Q}}$};
      \node at ($(v2.south)+(0,-0.2cm)$) [anchor=north] {$\mathrlap{\gen{P}}\phantom{\gen{Q}}$};
      \node at ($(v3.south)+(0,-0.2cm)$) [anchor=north] {$\mathrlap{\dot{\gen{S}}_1}\phantom{\gen{Q}}$};
    \end{tikzpicture}
  }
  
  \caption{Two Dynkin diagrams for $\algPSU(1,1|2)$ with the simple roots indicated.}
  \label{fig:dynkin-su22}
\end{figure}

In a superalgebra the choice of simple roots and corresponding Dynkin diagram is not unique. Here we will mainly consider two different gradings of $\algPSU(1,1|2)$. In the first one, which we refer to as the $\algSU(2)$ grading, the simple roots are given by
\begin{equation}
  \gen{Q}_2 , \qquad \gen{J}^2{}_1 , \qquad \dot{\gen{Q}}^1 ,
\end{equation}
while the $\algSL(2)$ grading, the simple roots are given by
\begin{equation}
  \gen{S}^2 , \qquad \gen{P} , \qquad \dot{\gen{S}}_1 .
\end{equation}
The corresponding Dynkin diagrams are shown in figure~\ref{fig:dynkin-su22}.

\paragraph{Representations.}

In this paper we are interested in unitary highest weight representations of $\algPSU(1,1|2)$. Such a representation can be parametrised by the charges $h$ and $j$ that the highest weight state\footnote{%
  The assignment of weights to the states of the representation depends on the choice of grading, as discussed above. For concreteness we will always write the weights corresponding to the $\algSU(2)$ grading.%
} %
carries under the Cartan elements $\gen{D}$ and $\gen{J}^1{}_1$, and by the eigenvalue $b$ under the Cartan element of the $\algSU(2)_{\bullet}$ automorphism. We will denote a generic such module by $(h;j)_b$. The representation can be decomposed into a number of irreducible representations of the bosonic subalgebra $\algSU(1,1) \oplus \algSU(2)$. Since four of the eight supercharges act as creation operators on the highest weight state we in general obtain sixteen such sub-modules.

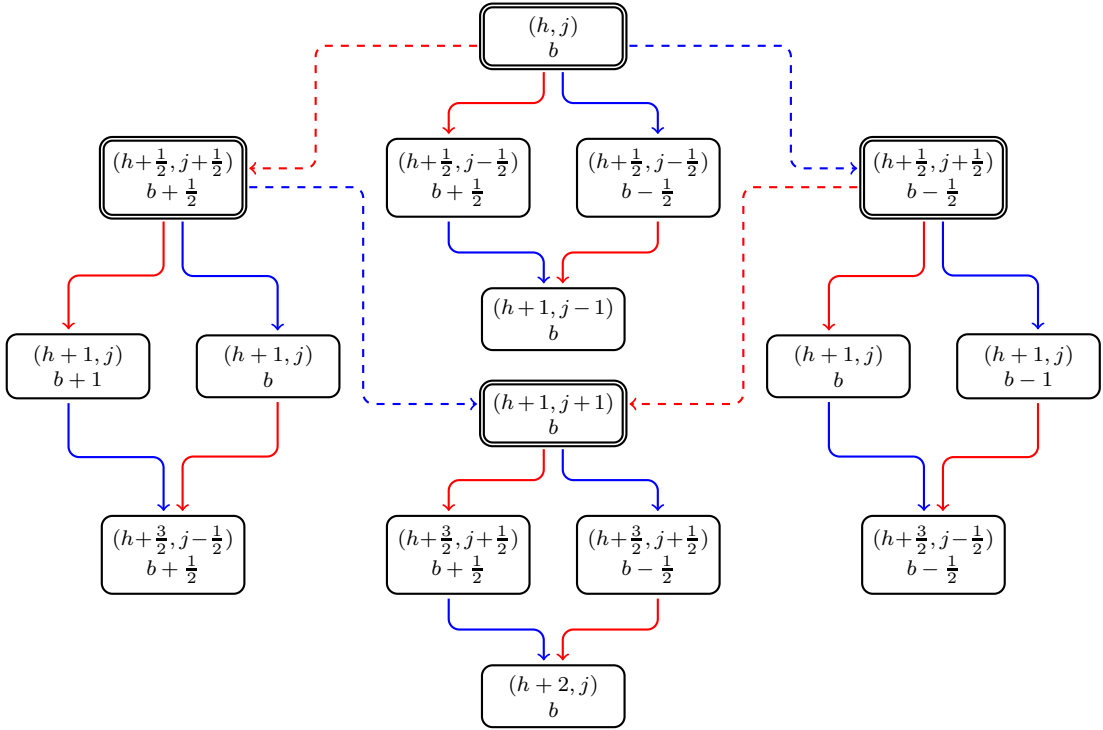
\begin{figure}
  \newlength{\vertdist}
  \setlength\vertdist{1.25cm}

  \centering
  \begin{tikzpicture}[
    ]

    \node (v11) at (0,0) [hw box] {\parbox{1.6cm}{\centering\scriptsize $(h,j)$\\$b$}};

    \node (v21) at (-5.0,-1.5\vertdist) [hw box] {\parbox{1.6cm}{\centering\scriptsize $(h+\tfrac{1}{2},j+\tfrac{1}{2})$\\$b+\tfrac{1}{2}$}};
    \node (v22) at (-1.25,-1.5\vertdist) [box] {\parbox{1.6cm}{\centering\scriptsize $(h+\tfrac{1}{2},j-\tfrac{1}{2})$\\${b+\tfrac{1}{2}}$}};
    \node (v23) at (+1.25,-1.5\vertdist) [box] {\parbox{1.6cm}{\centering\scriptsize $(h+\tfrac{1}{2},j-\tfrac{1}{2})$\\${b-\tfrac{1}{2}}$}};
    \node (v24) at (+5.0,-1.5\vertdist) [hw box] {\parbox{1.6cm}{\centering\scriptsize $(h+\tfrac{1}{2},j+\tfrac{1}{2})$\\${b-\tfrac{1}{2}}$}};

    \node (v31) at (-6.25,-3.5\vertdist) [box] {\parbox{1.6cm}{\centering\scriptsize $(h+1,j)$\\${b+1}$}};
    \node (v32) at (-3.75,-3.5\vertdist) [box] {\parbox{1.6cm}{\centering\scriptsize $(h+1,j)$\\${b}$}};
    \node (v33) at ( 0.0,-3\vertdist) [box] {\parbox{1.6cm}{\centering\scriptsize $(h+1,j-1)$\\${b}$}};
    \node (v34) at ( 0.0,-4\vertdist) [hw box] {\parbox{1.6cm}{\centering\scriptsize $(h+1,j+1)$\\${b}$}};
    \node (v35) at (+3.75,-3.5\vertdist) [box] {\parbox{1.6cm}{\centering\scriptsize $(h+1,j)$\\${b}$}};
    \node (v36) at (+6.25,-3.5\vertdist) [box] {\parbox{1.6cm}{\centering\scriptsize $(h+1,j)$\\${b-1}$}};

    \node (v41) at (-5.0,-5.5\vertdist) [box] {\parbox{1.6cm}{\centering\scriptsize $(h+\tfrac{3}{2},j-\tfrac{1}{2})$\\${b+\tfrac{1}{2}}$}};
    \node (v42) at (-1.25,-5.5\vertdist) [box] {\parbox{1.6cm}{\centering\scriptsize $(h+\tfrac{3}{2},j+\tfrac{1}{2})$\\${b+\tfrac{1}{2}}$}};
    \node (v43) at (+1.25,-5.5\vertdist) [box] {\parbox{1.6cm}{\centering\scriptsize $(h+\tfrac{3}{2},j+\tfrac{1}{2})$\\${b-\tfrac{1}{2}}$}};
    \node (v44) at (+5.0,-5.5\vertdist) [box] {\parbox{1.6cm}{\centering\scriptsize $(h+\tfrac{3}{2},j-\tfrac{1}{2})$\\${b-\tfrac{1}{2}}$}};

    \node (v51) at (0,-7\vertdist) [box] {\parbox{1.6cm}{\centering\scriptsize $(h+2,j)$\\$b$}};

    \draw [Q,->] ($(v11.south)+(-0.125,-0.05)$) |- ($(v11.south west)!0.5!(v22.north)$) -| ($(v22.north)+(-0.125,+0.05)$);
    \draw [QQ,->] ($(v11.south)+(+0.125,-0.05)$) |- ($(v11.south)!0.5!(v23.north)$) -| ($(v23.north)+(+0.125,+0.05)$);
    \draw [QQ,->] ($(v22.south)+(-0.125,-0.05)$) |- ($(v22.south)!0.5!(v33.north)$) -| ($(v33.north)+(-0.125,+0.05)$);
    \draw [Q,->] ($(v23.south)+(+0.125,-0.05)$) |- ($(v23.south)!0.5!(v33.north)$) -| ($(v33.north)+(+0.125,+0.05)$);

    \draw [Q,->] ($(v21.south)+(-0.125,-0.05)$) |- ($(v21.south)!0.5!(v31.north)$) -| ($(v31.north)+(-0.125,+0.05)$);
    \draw [QQ,->] ($(v21.south)+(+0.125,-0.05)$) |- ($(v21.south)!0.5!(v32.north)$) -| ($(v32.north)+(+0.125,+0.05)$);
    \draw [QQ,->] ($(v31.south)+(-0.125,-0.05)$) |- ($(v31.south)!0.5!(v41.north)$) -| ($(v41.north)+(-0.125,+0.05)$);
    \draw [Q,->] ($(v32.south)+(+0.125,-0.05)$) |- ($(v32.south)!0.5!(v41.north)$) -| ($(v41.north)+(+0.125,+0.05)$);

    \draw [Q,->] ($(v24.south)+(-0.125,-0.05)$) |- ($(v24.south)!0.5!(v35.north)$) -| ($(v35.north)+(-0.125,+0.05)$);
    \draw [QQ,->] ($(v24.south)+(+0.125,-0.05)$) |- ($(v24.south)!0.5!(v36.north)$) -| ($(v36.north)+(+0.125,+0.05)$);
    \draw [QQ,->] ($(v35.south)+(-0.125,-0.05)$) |- ($(v35.south)!0.5!(v44.north)$) -| ($(v44.north)+(-0.125,+0.05)$);
    \draw [Q,->] ($(v36.south)+(+0.125,-0.05)$) |- ($(v36.south)!0.5!(v44.north)$) -| ($(v44.north)+(+0.125,+0.05)$);

    \draw [Q,->] ($(v34.south)+(-0.125,-0.05)$) |- ($(v34.south)!0.5!(v42.north)$) -| ($(v42.north)+(-0.125,+0.05)$);
    \draw [QQ,->] ($(v34.south)+(+0.125,-0.05)$) |- ($(v34.south)!0.5!(v43.north)$) -|  ($(v43.north)+(+0.125,+0.05)$);
    \draw [QQ,->] ($(v42.south)+(-0.125,-0.05)$) |- ($(v42.south)!0.5!(v51.north)$) -| ($(v51.north)+(-0.125,+0.05)$);
    \draw [Q,->] ($(v43.south)+(+0.125,-0.05)$) |- ($(v43.south)!0.5!(v51.north)$) -| ($(v51.north)+(+0.125,+0.05)$);

    \draw [Q,dashed,->] ($(v11.west)+(-0.05,-0.125)$) -| ($(v21.north)!0.5!(v22.north)$) |- ($(v21.east)+(+0.05,+0.125)$);
    \draw [QQ,dashed,->] ($(v11.east)+(+0.05,-0.125)$) -| ($(v23.north)!0.5!(v24.north)$) |- ($(v24.west)+(-0.05,+0.125)$);

    \draw [Q,dashed,->] ($(v24.west)+(-0.05,-0.125)$) -| ($(v34.north)!0.5!(v24.south)$) |- ($(v34.east)+(+0.05,+0.125)$);
    \draw [QQ,dashed,->] ($(v21.east)+(+0.05,-0.125)$) -| ($(v34.north)!0.5!(v21.south)$) |- ($(v34.west)+(-0.05,+0.125)$);

  \end{tikzpicture}

  \caption{A generic $\algPSU(1,1|2)$ representation $[h,j]_b$ contains sixteen irreducible $\algSU(1,1) \oplus \algSU(2)$ submodules, which are here represented by their charges under $\algSU(1,1) \oplus \algSU(2)$ and under $\algSU(2)_{\bullet}$. In this figure the charges of these submodules are depicted together with their eigenvalues under $\gen{J}_{\bullet}$. The solid red (blue) arrows indicate the action of the generators $\gen{Q}_1$ ($\dot{\gen{Q}}^2$) and the dashed red (blue) arrows indicate the action of generators $\gen{Q}_2$ ($\dot{\gen{Q}}^1$). Note that not all such actions are depicted. For $h=j$ the representation $(j,j)_b$ becomes reducible and splits into four short representations with highest weight states corresponding to the states in the double boxes.}
  \label{fig:long-irrep}
\end{figure}

\section{Computation of the modified elliptic genus}
\label{app:elliptic-genus}

In this appendix we compute the elliptic genus for the spin chain associated to $\AdS_3\times \Sphere^3\times \Torus^4$, defined in equation \eqref{eq:megspinchain}. We deal with the zero mode insertion by first considering the ``partition function''
\begin{equation}
  Z(Q,q,y,\bar{y}) = \mathrm{Tr}_{|\textrm{anything}\rangle\sL \otimes |\textrm{chiral primary}\rangle\sR} (-1)^F Q^d\, q^{2 \gen{D}\sL} y^{2 \gen{J}\sL} \bar{y}^{2\gen{J}\sR} ,
\end{equation}
from which the modified elliptic genus can be computed using\footnote{We remind the reader that the elliptic genus is typically defined in the Ramond sector, where it receives contributions only from the right-moving Ramond ground states (see footnote \ref{ftn:egramond}). Ground states are then mapped under spectral flow to the highest-weight component of chiral primary multiplets in the NS sector, that is states such that $D\sR = J\sR$.}
\begin{equation}
  \label{eq:egasder}
  \tilde{\mathcal{E}}_{2}(Q,q,y) = \bar{y} \partial_{\bar{y}} (\bar{y} \partial_{\bar{y}} Z(Q,q,y,\bar{y}))|_{\bar{y}=1} .
\end{equation} 
The quantum number $d$ is the degree introduced in section~\ref{sec:fermioniczeromodes}, following~\cite{deBoer:1998us,Maldacena:1999bp}.

We first consider the partition function in the single-trace sector $Z_{\mathrm{s.t.}}(Q,q,y,\bar{y})$.
By looking at tables \ref{tab:massive-fields} and \ref{tab:massless-fields} we see that the fields that can be used to build 1/4-BPS states are
\begin{equation}
  \label{eq:1/4-BPS-ingr}
  \phi^{+ \dot{\alpha}}~, \quad \psi_{\sL}^{+ \dot{a}}~, \quad \chi_{\sL}^{\dot{\alpha}a}~, \quad \chi_{\sR}^{+a} ,
\end{equation}
and we can also act with an arbitrary number of left derivatives. Consider a single-trace operator $\mathcal{O} = \tr A$ of degree $d$, (left) conformal dimension $D\sL$, and $R$-charge $(J\sL,J\sR)$. Without loss of generality, we consider the case where $A$ does not contain $\chi_{\sR}^{+\pm}$. Then the spectrum necessarily contains the four operators
\begin{align}
  & \tr A~,  & & \tr (A \chi_R^{++})~, & & \tr (A \chi_R^{+-})~, & & \tr (A \chi_R^{++} \chi_R^{+-}) .
\end{align}
It is evident that these operators have the same $D\sL$, $J\sL$ and $d$ quantum numbers. Therefore, they contribute to the partition function as
\begin{equation}
  \label{eq:stPF}
  Z_{\mathrm{s.t.}}(Q,q,y,\bar{y}) = \ldots + Q^d q^{2D\sL} y^{2J\sL} \bar{y}^{2J\sR} \left(\bar{y} - 2 + \frac{1}{\bar{y}}\right) + \ldots ,
\end{equation}
where we have taken into account the effect of spectral flow on the right sector.\footnote{Spectral flow is implemented in a sector of a given degree $d$ with an ``effective'' central charge equal to $6d$, as in \cite{Maldacena:1999bp}. Of course this also changes the ``overall'' charge $J\sR$ in \eqref{eq:stPF}, but we will see that, thanks to the double logarithmic derivative in \eqref{eq:egasder}, the elliptic genus is independent of overall powers of $\bar{y}$ in the partition function.}  Given a single-trace partition function of the form
\begin{equation}
  Z_{\mathrm{s.t.}}(Q,q,y,\bar{y}) = \sum_{m,n,j,k} c(m,n,j,k) Q^m q^n y^j \bar{y}^k ,
\end{equation}
the full partition function including multi-trace contributions is given by
\begin{equation}
  Z(Q,q,y,\bar{y}) = \prod_{m,n,j,k} \frac{1}{(1-Q^m q^n y^j \bar{y}^k)^{c(m,n,j,k)}} .
\end{equation}
In summary, we have shown that
\begin{equation}
  Z(Q,q,y,\bar{y}) = \sideset{}{'}{\prod}_{m,n,j,k} \frac{(1-Q^m q^n y^j \bar{y}^k)^2}{(1-Q^m q^n y^j \bar{y}^{k+1})(1-Q^m q^n y^j \bar{y}^{k-1})} ,
\end{equation}
where the prime on the product symbol indicates that we multiply over the $(m,n,j,k)$ corresponding to the charges of all the possible single-trace operators $\tr A$ with no $\chi_{\sR}^{+\pm}$ insertions. Taking the double derivative and setting $\bar{y}=1$ as in \eqref{eq:egasder}, we obtain
\begin{equation}
  \label{eq:genegT4}
  \tilde{\mathcal{E}}_{2}(Q,q,y) = \sideset{}{'}{\sum}_{m,n,j,k} \frac{2 Q^m q^n y^j}{(1-Q^m q^n y^j)^2} .
\end{equation}
Since we are interested in the modified elliptic genus only up to powers of $q$ such that $D\sL < d/4$ \cite{deBoer:1998us,Maldacena:1999bp}, we see from the previous formula that we need to consider only \emph{single-trace} operators such that $D\sL < d/4$. All the fields in \eqref{eq:1/4-BPS-ingr} except $\chi_{\sR}$ satisfy $D\sL \geq d/2$, so a single-trace operator $\mathcal{O}_k$ that consists of a string of $k$ operators not involving $\chi_{\sR}$ satisfies
\begin{equation}
  D\sL \geq \frac{k}{2}~, \qquad d \leq k + 1 ,
\end{equation}
which together with $D\sL < d/4$ implies $k < 1$. This means that the only single-trace operators that satisfy our criterion are
\begin{align}
  & \tr 1~,  & & \tr \chi_{\sR}^{++}~, & & \tr \chi_{\sR}^{+-}~, & & \tr \chi_{\sR}^{++} \chi_{\sR}^{+-}~.
\end{align}
As a consequence, from \eqref{eq:genegT4} we obtain
\begin{equation}
\tilde{\mathcal{E}}_2(Q,q,y) = \frac{2Q}{(1-Q)^2} + \ldots = \sum_{N} 2 N\, Q^N +\ldots ,
\end{equation}
where the ellipses denote terms that correspond to states with large conformal dimension compared to $N$. This reproduces the modified elliptic genus of \cite{Maldacena:1999bp}.

\section{Multiplet joining and length-changing effects}
We have seen in section~\ref{sec:spinchain} that when states that saturate the BPS bound~\eqref{eq:bpsconditionSmall} receive an anomalous dimension, we expect short multiplets to join into long ones.  Below we detail the action of the $\algPSU(1,1|2)_{\sL}\oplus\algPSU(1,1|2)_{\sR}$ supercharges on the spin-chain states on such multiplets.

\subsection{Massive excitations}
\label{app:massive-joining}
Let us first consider the case of the accidentally-short multiplet with highest-weight state given by~\eqref{eq:state-su2L}.
We now need to understand how to obtain states with the charges given in~\eqref{eq:massive-L-multiplet-splitting}. 
The left-moving supercharges $\gen{Q}_{\sL\,2}$ and $\dot{\gen{Q}}_{\sL}^1$ schematically act on it as on a standard long multiplet, as illustrated in the following figure\footnote{%
  See also figure~\ref{fig:long-irrep} in appendix~\ref{sec:psu112} for a more complete illustration of a long $\algPSU(1,1|2)$ multiplet.%
} %
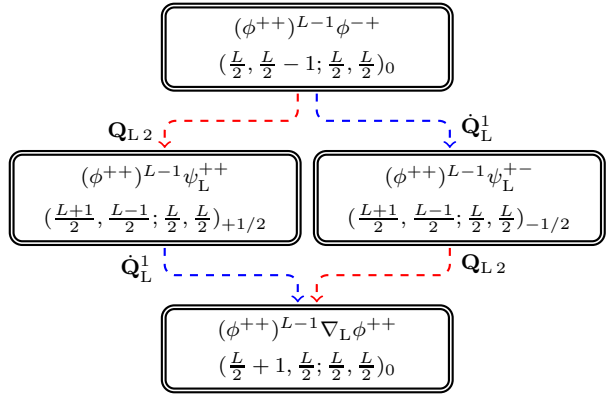
\begin{equation*}
  \begin{tikzpicture}
    \node (a) at (0,+2) [hw box] {\parbox{3.5cm}{\centering\scriptsize $(\phi^{++})^{L-1} \phi^{-+}$ \\[4pt] $( \tfrac{L}{2},\tfrac{L}{2}-1;\tfrac{L}{2},\tfrac{L}{2} )_0$}};
    \node (b) at (-2,0) [hw box] {\parbox{3.5cm}{\centering\scriptsize $(\phi^{++})^{L-1} \psi_{\sL}^{++}$ \\[4pt] $(\tfrac{L+1}{2},\tfrac{L-1}{2};\tfrac{L}{2},\tfrac{L}{2})_{+1/2}$}};
    \node (c) at (+2,0) [hw box] {\parbox{3.5cm}{\centering\scriptsize $(\phi^{++})^{L-1} \psi_{\sL}^{+-}$ \\[4pt] $(\tfrac{L+1}{2},\tfrac{L-1}{2};\tfrac{L}{2},\tfrac{L}{2})_{-1/2}$}};
    \node (d) at (0,-2) [hw box] {\parbox{3.5cm}{\centering\scriptsize $(\phi^{++})^{L-1} \nabla_{\sL} \phi^{++}$ \\[4pt] $(\tfrac{L}{2}+1,\tfrac{L}{2};\tfrac{L}{2},\tfrac{L}{2})_0$}};
    
    \draw [Q,dashed,->] ($(a.south)+(-0.125,-0.05)$) |- ($(a.south)!0.5!(b.north)$) -| ($(b.north)+(+0.125,+0.05)$);
    \draw [QQ,dashed,->] ($(a.south)+(+0.125,-0.05)$) |- ($(a.south)!0.5!(c.north)$) -| ($(c.north)+(-0.125,+0.05)$);
      
    \draw [QQ,dashed,->] ($(b.south)+(+0.125,-0.05)$) |- ($(b.south)!0.5!(d.north)$) -| ($(d.north)+(-0.125,+0.05)$);
    \draw [Q,dashed,->] ($(c.south)+(-0.125,-0.05)$) |- ($(c.south)!0.5!(d.north)$) -| ($(d.north)+(+0.125,+0.05)$);

    \node at ($(b.north)+(+0.125,0)$) [anchor=south east] {\scriptsize $\gen{Q}_{\sL\,2}$};
    \node at ($(b.south)+(+0.125,0)$) [anchor=north east] {\scriptsize $\dot{\gen{Q}}_{\sL}^1$};
    \node at ($(c.north)+(-0.125,0)$) [anchor=south west] {\scriptsize $\dot{\gen{Q}}_{\sL}^1$};
    \node at ($(c.south)+(-0.125,0)$) [anchor=north west] {\scriptsize $\gen{Q}_{\sL\,2}$};
  \end{tikzpicture}
\end{equation*}
However, the right-moving supercharges $\gen{Q}_{\sR\,2}$ and $\dot{\gen{Q}}_{\sR}^1$ to leading order annihilate the 1/4-BPS multiplet. For them to have a non-trivial action on the state~\eqref{eq:state-su2L} we need to let them insert extra sites into the spin-chain state,
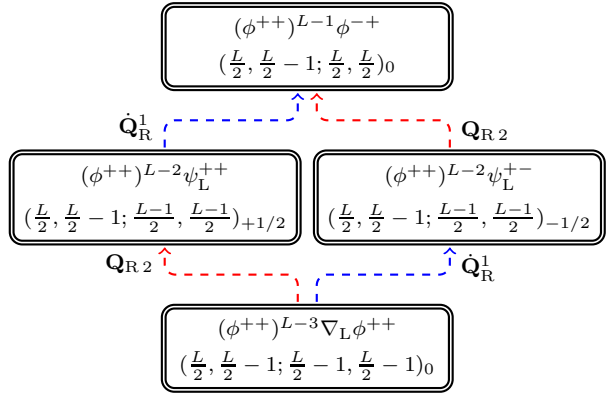
\begin{equation*}
  \begin{tikzpicture}
          \node (a) at (0,+2) [hw box] {\parbox{3.5cm}{\centering\scriptsize $(\phi^{++})^{L-1} \phi^{-+}$ \\[4pt] $(\tfrac{L}{2},\tfrac{L}{2}-1;\tfrac{L}{2},\tfrac{L}{2})_0$}};
      \node (b) at (-2,0) [hw box] {\parbox{3.5cm}{\centering\scriptsize $(\phi^{++})^{L-2} \psi_{\sL}^{++}$ \\[4pt] $(\tfrac{L}{2},\tfrac{L}{2}-1;\tfrac{L-1}{2},\tfrac{L-1}{2})_{+1/2}$}};
      \node (c) at (+2,0) [hw box] {\parbox{3.5cm}{\centering\scriptsize $(\phi^{++})^{L-2} \psi_{\sL}^{+-}$ \\[4pt] $(\tfrac{L}{2},\tfrac{L}{2}-1;\tfrac{L-1}{2},\tfrac{L-1}{2})_{-1/2}$}};
      \node (d) at (0,-2) [hw box] {\parbox{3.5cm}{\centering\scriptsize $(\phi^{++})^{L-3} \nabla_{\sL} \phi^{++}$ \\[4pt] $(\tfrac{L}{2},\tfrac{L}{2}-1;\tfrac{L}{2}-1,\tfrac{L}{2}-1)_0$}};

      \draw [QQ,dashed,->] ($(b.north)+(+0.125,+0.05)$) |- ($(a.south)!0.5!(b.north)$) -| ($(a.south)+(-0.125,-0.05)$);
      \draw [Q,dashed,->]  ($(c.north)+(-0.125,+0.05)$) |- ($(a.south)!0.5!(c.north)$) -| ($(a.south)+(+0.125,-0.05)$);
      
      \draw [Q,dashed,->]  ($(d.north)+(-0.125,+0.05)$) |- ($(b.south)!0.5!(d.north)$) -| ($(b.south)+(+0.125,-0.05)$);
      \draw [QQ,dashed,->] ($(d.north)+(+0.125,+0.05)$) |- ($(c.south)!0.5!(d.north)$) -| ($(c.south)+(-0.125,-0.05)$);

      \node at ($(b.north)+(+0.125,0)$) [anchor=south east] {\scriptsize $\dot{\gen{Q}}_{\sR}^1$};
      \node at ($(b.south)+(+0.125,0)$) [anchor=north east] {\scriptsize $\gen{Q}_{\sR\,2}$};
      \node at ($(c.north)+(-0.125,0)$) [anchor=south west] {\scriptsize $\gen{Q}_{\sR\,2}$};
      \node at ($(c.south)+(-0.125,0)$) [anchor=north west] {\scriptsize $\dot{\gen{Q}}_{\sR}^1$};
  \end{tikzpicture}
\end{equation*}
Note that the right-moving supercharges at each step increase the length of the state by one, and precisely yields the highest-weight states~\eqref{eq:massive-L-hws}.

Similarly, a highest weight state with a single right-moving excitation
\begin{equation}
  \ket{ (\phi^{++})^{L-1} \nabla_{\sR} \phi^{++} } + \text{permutations}
\end{equation}
has charges $(\tfrac{L}{2},\tfrac{L}{2} ; \tfrac{L}{2}+1,\tfrac{L}{2})_0$. This gives a short representation under the $\algPSU(1,1|2)_{\sL}$ algebra, but a long one under $\algPSU(1,1|2)_{\sR}$. For this representation to be deformed by an anomalous dimension it needs to join up with three other states with charges
\begin{equation}
  (\tfrac{L+1}{2},\tfrac{L+1}{2} ; \tfrac{L}{2}+1,\tfrac{L}{2})_{+\tfrac{1}{2}} , \quad
  (\tfrac{L+1}{2},\tfrac{L+1}{2} ; \tfrac{L}{2}+1,\tfrac{L}{2})_{-\tfrac{1}{2}} , \quad
  (\tfrac{L}{2}+1,\tfrac{L}{2}+1 ; \tfrac{L}{2}+1,\tfrac{L}{2})_{0} ,
\end{equation}
to form a long $\algPSU(1,1|2)_{\sL}$ representation. Again this is made possible by length-changing actions of the supercharges, which is summarised in the following figure
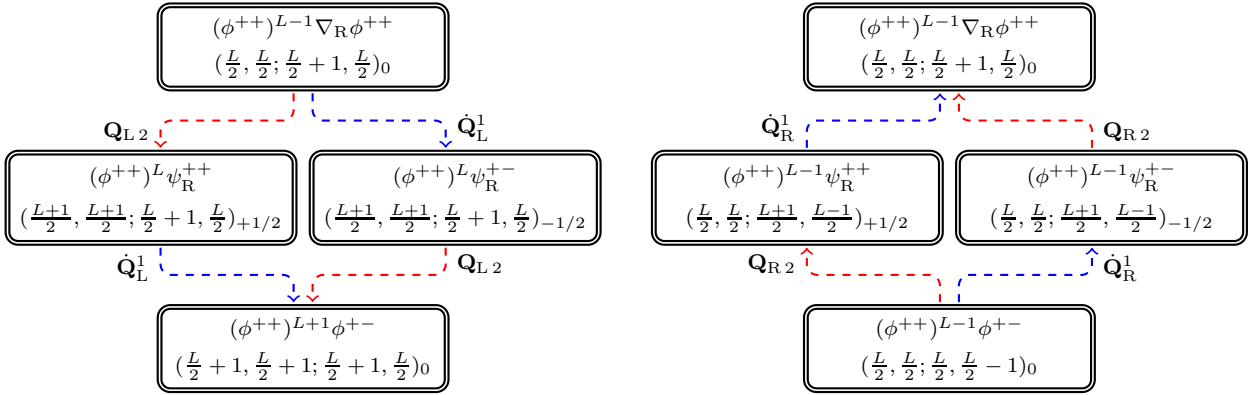
\begin{equation*}
  \begin{tikzpicture}
    \useasboundingbox (-4cm,+2.75cm) rectangle (+4cm,-2.75cm);

    \begin{scope}[xshift=-4.25cm]
      \node (a) at (0,+2) [hw box] {\parbox{3.5cm}{\centering\scriptsize $(\phi^{++})^{L-1} \nabla_{\sR} \phi^{++}$ \\[4pt] $(\tfrac{L}{2},\tfrac{L}{2};\tfrac{L}{2}+1,\tfrac{L}{2})_0$}};
      \node (b) at (-2,0) [hw box] {\parbox{3.5cm}{\centering\scriptsize $(\phi^{++})^{L} \psi_{\sR}^{++}$ \\[4pt] $(\tfrac{L+1}{2},\tfrac{L+1}{2};\tfrac{L}{2}+1,\tfrac{L}{2})_{+1/2}$}};
      \node (c) at (+2,0) [hw box] {\parbox{3.5cm}{\centering\scriptsize $(\phi^{++})^{L} \psi_{\sR}^{+-}$ \\[4pt] $(\tfrac{L+1}{2},\tfrac{L+1}{2};\tfrac{L}{2}+1,\tfrac{L}{2})_{-1/2}$}};
      \node (d) at (0,-2) [hw box] {\parbox{3.5cm}{\centering\scriptsize $(\phi^{++})^{L+1} \phi^{+-}$ \\[4pt] $(\tfrac{L}{2}+1,\tfrac{L}{2}+1;\tfrac{L}{2}+1,\tfrac{L}{2})_0$}};

      \draw [Q,dashed,->] ($(a.south)+(-0.125,-0.05)$) |- ($(a.south)!0.5!(b.north)$) -| ($(b.north)+(+0.125,+0.05)$);
      \draw [QQ,dashed,->] ($(a.south)+(+0.125,-0.05)$) |- ($(a.south)!0.5!(c.north)$) -| ($(c.north)+(-0.125,+0.05)$);
      
      \draw [QQ,dashed,->] ($(b.south)+(+0.125,-0.05)$) |- ($(b.south)!0.5!(d.north)$) -| ($(d.north)+(-0.125,+0.05)$);
      \draw [Q,dashed,->] ($(c.south)+(-0.125,-0.05)$) |- ($(c.south)!0.5!(d.north)$) -| ($(d.north)+(+0.125,+0.05)$);

      \node at ($(b.north)+(+0.125,0)$) [anchor=south east] {\scriptsize $\gen{Q}_{\sL\,2}$};
      \node at ($(b.south)+(+0.125,0)$) [anchor=north east] {\scriptsize $\dot{\gen{Q}}_{\sL}^1$};
      \node at ($(c.north)+(-0.125,0)$) [anchor=south west] {\scriptsize $\dot{\gen{Q}}_{\sL}^1$};
      \node at ($(c.south)+(-0.125,0)$) [anchor=north west] {\scriptsize $\gen{Q}_{\sL\,2}$};
    \end{scope}

    \begin{scope}[xshift=+4.25cm]
      \node (a) at (0,+2) [hw box] {\parbox{3.5cm}{\centering\scriptsize $(\phi^{++})^{L-1} \nabla_{\sR} \phi^{++}$ \\[4pt] $(\tfrac{L}{2},\tfrac{L}{2};\tfrac{L}{2}+1,\tfrac{L}{2})_0$}};
      \node (b) at (-2,0) [hw box] {\parbox{3.5cm}{\centering\scriptsize $(\phi^{++})^{L-1} \psi_{\sR}^{++}$ \\[4pt] $(\tfrac{L}{2},\tfrac{L}{2};\tfrac{L+1}{2},\tfrac{L-1}{2})_{+1/2}$}};
      \node (c) at (+2,0) [hw box] {\parbox{3.5cm}{\centering\scriptsize $(\phi^{++})^{L-1} \psi_{\sR}^{+-}$ \\[4pt] $(\tfrac{L}{2},\tfrac{L}{2};\tfrac{L+1}{2},\tfrac{L-1}{2})_{-1/2}$}};
      \node (d) at (0,-2) [hw box] {\parbox{3.5cm}{\centering\scriptsize $(\phi^{++})^{L-1} \phi^{+-}$ \\[4pt] $(\tfrac{L}{2},\tfrac{L}{2};\tfrac{L}{2},\tfrac{L}{2}-1)_0$}};

      \draw [QQ,dashed,->] ($(b.north)+(+0.125,+0.05)$) |- ($(a.south)!0.5!(b.north)$) -| ($(a.south)+(-0.125,-0.05)$);
      \draw [Q,dashed,->]  ($(c.north)+(-0.125,+0.05)$) |- ($(a.south)!0.5!(c.north)$) -| ($(a.south)+(+0.125,-0.05)$);
      
      \draw [Q,dashed,->]  ($(d.north)+(-0.125,+0.05)$) |- ($(b.south)!0.5!(d.north)$) -| ($(b.south)+(+0.125,-0.05)$);
      \draw [QQ,dashed,->] ($(d.north)+(+0.125,+0.05)$) |- ($(c.south)!0.5!(d.north)$) -| ($(c.south)+(-0.125,-0.05)$);

      \node at ($(b.north)+(+0.125,0)$) [anchor=south east] {\scriptsize $\dot{\gen{Q}}_{\sR}^1$};
      \node at ($(b.south)+(+0.125,0)$) [anchor=north east] {\scriptsize $\gen{Q}_{\sR\,2}$};
      \node at ($(c.north)+(-0.125,0)$) [anchor=south west] {\scriptsize $\gen{Q}_{\sR\,2}$};
      \node at ($(c.south)+(-0.125,0)$) [anchor=north west] {\scriptsize $\dot{\gen{Q}}_{\sR}^1$};
    \end{scope}
  \end{tikzpicture}
\end{equation*}
Here the left-moving supercharges increase the length by one at each step. Note however that these length-changing effects are not visible in the leading order Bethe equations, but only show up at higher orders in the coupling constant $h$.

\subsection{Massless excitations}
\label{app:massless-joining}
 
Let us consider the multiplet identified by~\eqref{eq:single-massless-magnon}. If the excitation has non-vanishing momentum, we expect the short multiplets to join and the supercharges to act according to the following diagrams:
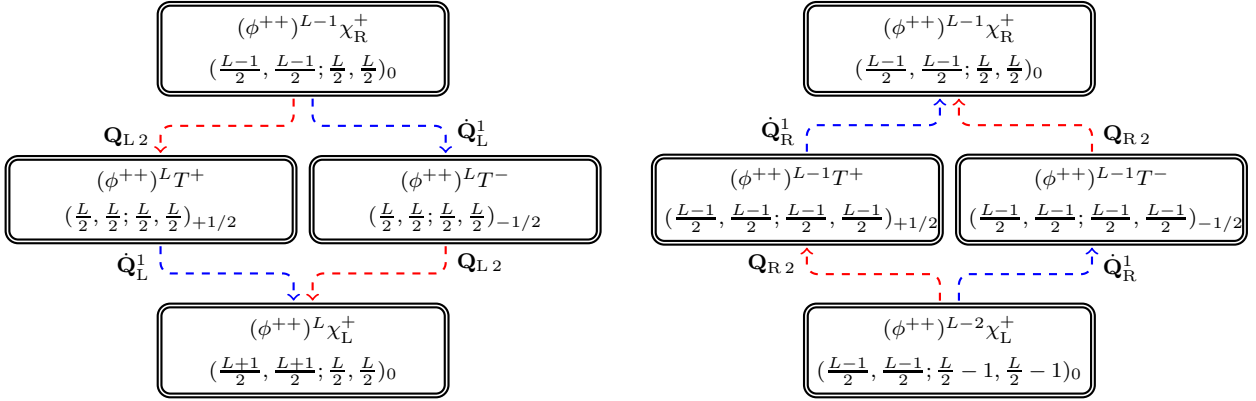
\begin{equation*}
  \begin{tikzpicture}
    \useasboundingbox (-4cm,+2.75cm) rectangle (+4cm,-2.75cm);
    \setlength\overfullrule{0pt}

    \begin{scope}[xshift=-4.25cm]
      \node (a) at (0,+2) [hw box] {\parbox{3.5cm}{\centering\scriptsize $(\phi^{++})^{L-1} \chi_{\sR}^+$ \\[4pt] $(\tfrac{L-1}{2},\tfrac{L-1}{2};\tfrac{L}{2},\tfrac{L}{2})_0$}};
      \node (b) at (-2,0) [hw box] {\parbox{3.5cm}{\centering\scriptsize $(\phi^{++})^{L} T^+$ \\[4pt] $(\tfrac{L}{2},\tfrac{L}{2};\tfrac{L}{2},\tfrac{L}{2})_{+1/2}$}};
      \node (c) at (+2,0) [hw box] {\parbox{3.5cm}{\centering\scriptsize $(\phi^{++})^{L} T^-$ \\[4pt] $(\tfrac{L}{2},\tfrac{L}{2};\tfrac{L}{2},\tfrac{L}{2})_{-1/2}$}};
      \node (d) at (0,-2) [hw box] {\parbox{3.5cm}{\centering\scriptsize $(\phi^{++})^{L} \chi_{\sL}^+$ \\[4pt] $(\tfrac{L+1}{2},\tfrac{L+1}{2};\tfrac{L}{2},\tfrac{L}{2})_0$}};

      \draw [Q,dashed,->] ($(a.south)+(-0.125,-0.05)$) |- ($(a.south)!0.5!(b.north)$) -| ($(b.north)+(+0.125,+0.05)$);
      \draw [QQ,dashed,->] ($(a.south)+(+0.125,-0.05)$) |- ($(a.south)!0.5!(c.north)$) -| ($(c.north)+(-0.125,+0.05)$);
      
      \draw [QQ,dashed,->] ($(b.south)+(+0.125,-0.05)$) |- ($(b.south)!0.5!(d.north)$) -| ($(d.north)+(-0.125,+0.05)$);
      \draw [Q,dashed,->] ($(c.south)+(-0.125,-0.05)$) |- ($(c.south)!0.5!(d.north)$) -| ($(d.north)+(+0.125,+0.05)$);

      \node at ($(b.north)+(+0.125,0)$) [anchor=south east] {\scriptsize $\gen{Q}_{\sL\,2}$};
      \node at ($(b.south)+(+0.125,0)$) [anchor=north east] {\scriptsize $\dot{\gen{Q}}_{\sL}^1$};
      \node at ($(c.north)+(-0.125,0)$) [anchor=south west] {\scriptsize $\dot{\gen{Q}}_{\sL}^1$};
      \node at ($(c.south)+(-0.125,0)$) [anchor=north west] {\scriptsize $\gen{Q}_{\sL\,2}$};
    \end{scope}

    \begin{scope}[xshift=+4.25cm]
      \node (a) at (0,+2) [hw box] {\parbox{3.5cm}{\centering\scriptsize $(\phi^{++})^{L-1} \chi_{\sR}^+$ \\[4pt] $(\tfrac{L-1}{2},\tfrac{L-1}{2};\tfrac{L}{2},\tfrac{L}{2})_0$}};
      \node (b) at (-2,0) [hw box] {\parbox{3.5cm}{\centering\scriptsize $(\phi^{++})^{L-1} T^+$ \\[4pt] $(\tfrac{L-1}{2},\tfrac{L-1}{2};\tfrac{L-1}{2},\tfrac{L-1}{2})_{+1/2}$}};
      \node (c) at (+2,0) [hw box] {\parbox{3.5cm}{\centering\scriptsize $(\phi^{++})^{L-1} T^-$ \\[4pt] $(\tfrac{L-1}{2},\tfrac{L-1}{2};\tfrac{L-1}{2},\tfrac{L-1}{2})_{-1/2}$}};
      \node (d) at (0,-2) [hw box] {\parbox{3.5cm}{\centering\scriptsize $(\phi^{++})^{L-2} \chi_{\sL}^+$ \\[4pt] $(\tfrac{L-1}{2},\tfrac{L-1}{2};\tfrac{L}{2}-1,\tfrac{L}{2}-1)_0$}};

      \draw [QQ,dashed,->] ($(b.north)+(+0.125,+0.05)$) |- ($(a.south)!0.5!(b.north)$) -| ($(a.south)+(-0.125,-0.05)$);
      \draw [Q,dashed,->]  ($(c.north)+(-0.125,+0.05)$) |- ($(a.south)!0.5!(c.north)$) -| ($(a.south)+(+0.125,-0.05)$);
      
      \draw [Q,dashed,->]  ($(d.north)+(-0.125,+0.05)$) |- ($(b.south)!0.5!(d.north)$) -| ($(b.south)+(+0.125,-0.05)$);
      \draw [QQ,dashed,->] ($(d.north)+(+0.125,+0.05)$) |- ($(c.south)!0.5!(d.north)$) -| ($(c.south)+(-0.125,-0.05)$);

      \node at ($(b.north)+(+0.125,0)$) [anchor=south east] {\scriptsize $\dot{\gen{Q}}_{\sR}^1$};
      \node at ($(b.south)+(+0.125,0)$) [anchor=north east] {\scriptsize $\gen{Q}_{\sR\,2}$};
      \node at ($(c.north)+(-0.125,0)$) [anchor=south west] {\scriptsize $\gen{Q}_{\sR\,2}$};
      \node at ($(c.south)+(-0.125,0)$) [anchor=north west] {\scriptsize $\dot{\gen{Q}}_{\sR}^1$};
    \end{scope}

  \end{tikzpicture}
\end{equation*}
Note that neither the left nor the right supercharges now preserve the length of the spin chain state. The structure of the dynamic spin chain involving massless modes is further discussed in appendix~\ref{app:dynamic-sc}.

As we have just seen, in order to fill out the full $\algPSU(1,1|2)_{\sL} \oplus \algPSU(1,1|2)_{\sR}$ long multiplet we need four states containing a single massless excitation. Let us see how to obtain these from the Bethe equations. We already know that we can obtain 
\begin{equation}
  \ket{ (\phi^{++})^{L-1} \chi_{\sR}^{+\pm} } + \text{permutations}
\end{equation}
from a solution with excitation number $N_0$. As remarked this state is 1/2-BPS at $h=0$. This means that it is annihilated by the $\algPSU(1,1|2)_{\sL} \oplus \algPSU(1,1|2)_{\sR}$ supercharges corresponding to the massive auxiliary Bethe roots $v_{1,k}$, $v_{3,k}$, $v_{\bar{1},k}$ and $v_{\bar{3},k}$. The only remaining possibility for constructing massless single particle states is hence to turn on the massless auxiliary roots $r_{1,k}$ and $r_{3,k}$. By reading off the corresponding charges from equation~\eqref{eq:charges} we can identify the following configurations
\begin{center}
  \begin{tabular}{lccc}
    \toprule
    State & $N_0$ & $N_1$ & $N_3$ \\
    \midrule
    $(\phi^{++})^{L-1} \chi_{\sR}^{+\pm}$ & $1$ & $0$ & $0$ \\
    $(\phi^{++})^{L\phantom{-1}} T^{+\pm}$ & $1$ & $1$ & $0$ \\
    $(\phi^{++})^{L\phantom{-1}} T^{-\pm}$ & $1$ & $0$ & $1$ \\
    $(\phi^{++})^{L\phantom{-1}} \chi_{\sL}^{+\pm}$ & $1$ & $1$ & $1$ \\
    \bottomrule
  \end{tabular}
\end{center}
As discussed, the $\algSU(2)_{\circ}$ index on these excitations is not directly encoded in the Bethe roots but needs to be kept track of externally.
Note that in the above spin-chain description the left and right parts of the spin-chain sites enter in a non-symmetric fashion. This is because of our choice of grading. By performing a set of fermionic dualities we can change the grading in such a way that the role of the left and right copies of $\algPSU(1,1|2)$ are exchanged~\cite{Borsato:2014hja}.

So far we have described the fundamental excitations of the spin chain. However, the system also contains additional fields such as the fermion $\chi_{\sL}^{-+}$. In the Bethe equations, a state containing this field is a \emph{multi-excitation} state. It can be obtained from a state containing the field $\chi_{\sL}^{++}$ by further turning on a massive momentum-carrying root $u$. This leads to the mixing of the states
\begin{equation}
  \ket{ (\phi^{++})^L \chi_{\sL}^{-\pm} } + \text{permutations}
\end{equation}
and
\begin{equation}
  \ket{ (\phi^{++})^{L-1} \phi^{-+} \chi_{\sL}^{+\pm} } + \text{permutations} .
\end{equation}
Similarly there is a mixing between the states
\begin{equation}
  \ket{ (\phi^{++})^{L-1} \nabla_{\sR} T^{+\pm} } + \text{permutations}
\end{equation}
and
\begin{equation}
  \ket{ (\phi^{++})^{L-2} \psi_{\sR}^{++} \chi_{\sR}^{+\pm} } + \text{permutations}
\end{equation}
which both have excitation numbers $M_{\bar{2}} = M_{\bar{1}} = N_0 = 1$.\footnote{%
  Note that at weak coupling these states do \emph{not} mix with
  \begin{equation*}
    \ket{ (\phi^{++})^{L-1} \nabla_{\sR} \phi^{++} T^{+\pm} } + \text{permutations},
  \end{equation*}
  which has $M_{\bar{2}} = N_0 = N_1 = 1$.
} %

\section{Dynamic spin chains}
\label{app:dynamic-sc}

As discussed above, several short multiplets can combine into a long multiplet through length-changing effects. In order to see this more explicitly we can restrict the spin-chain to include only massive excitations, so that the spin chain becomes homogenous with all the sites transforming in the same $\algPSU(1,1|2) \oplus \algPSU(1,1|2)$ representation. We furthermore consider a closed subsector consisting of states that saturate the bound
\begin{equation}
  D \geq 2J_{\sL} + J_{\bullet} ,
\end{equation}
which means that the states can be built from the three fields
\begin{equation}
  \phi^+ \equiv \phi^{++} , \quad
  \phi^- \equiv \phi^{+-} , \quad
  \psi \equiv \psi_{\sR}^{++} .
\end{equation}
The sector is preserved by the algebra $\algSU(1|1)_{\sL} \oplus \algSU(1|2)_{\sR}$ which is generated by the supercharges
\begin{equation}
  \dot{\gen{Q}} \equiv \dot{\gen{Q}}_{\sL}^1 , \qquad
  \dot{\gen{S}} \equiv \dot{\gen{S}}^{\sL}_1 , \qquad
  \gen{Q}_{\alpha} \equiv \gen{Q}^{\sR}_{\alpha} , \qquad
  \gen{S}^{\alpha} \equiv \gen{S}_{\sR}^{\alpha} ,
\end{equation}
together with the bosonic generators
\begin{equation}
  \gen{J}^{\alpha}{}_{\beta} \equiv (\gen{J}_{\sR})^{\alpha}{}_{\beta} , \qquad
  \gen{L} = 2\gen{J}_{\sL} , \qquad
  \gen{J}_{\bullet} , \qquad
  \gen{D} .
\end{equation}
The generator $\gen{L}$ measures the length of the spin chain, which is a preserved quantity in this sector.\footnote{%
  As we will see below, the spin-chain length $\gen{L}$ commutes with the spin-chain Hamiltonian. However, the supercharges $\dot{\gen{Q}}$ and $\dot{\gen{S}}$ change the length and hence the full symmetry algebra acts on a dynamic spin chain.%
} %
We further note that in this sector
\begin{equation}
  \gen{D}_{\sL} = \tfrac{1}{2} ( \gen{D} - \gen{J}_{\bullet} ) , \qquad
  \gen{D}_{\sR} = \tfrac{1}{2} ( \gen{D} + \gen{J}_{\bullet} ) .
\end{equation}
The algebra then takes the form
\begin{equation}
  \acomm{\dot{\gen{Q}}}{\dot{\gen{S}}} = \tfrac{1}{2} ( \gen{D} - \gen{J}_{\bullet} - \gen{L} ) , \qquad
  \acomm{\gen{Q}_{\alpha}}{\gen{S}^{\beta}} = \tfrac{1}{2} ( \gen{D} + \gen{J}_{\bullet} ) + \gen{J}^{\beta}{}_{\alpha} .
\end{equation}
The charges of the fields and generators under the various $\algU(1)$:s are shown in the following table
\begin{center}
  \begin{tabular}{ccccccc}
    \toprule
    & $D$ & $J_{\sR}$ & $J_{\bullet}$ & $\phantom{+}L$ & $D - L - J_{\bullet}$ \\
    \midrule
    $\phi^+$ & $+1$ & $+\tfrac{1}{2}$ & $\phantom{+}0$ & $\phantom{+}1$ & $0$ \\
    $\phi^-$ & $+1$ & $-\tfrac{1}{2}$ & $\phantom{+}0$ & $\phantom{+}1$ & $0$ \\
    $\psi$ & $+\tfrac{3}{2}$ & $\phantom{+}0$ & $+\tfrac{1}{2}$ & $\phantom{+}1$ & $0$ \\
    \midrule
    $\dot{\gen{Q}}$ & $+\tfrac{1}{2}$ & $\phantom{+}0$ & $-\tfrac{1}{2}$ & $+1$ & $0$ \\
    $\dot{\gen{S}}$ & $-\tfrac{1}{2}$ & $\phantom{+}0$ & $+\tfrac{1}{2}$ & $-1$ & $0$ \\
    \midrule
    $\gen{Q}_+$ & $+\tfrac{1}{2}$ & $-\tfrac{1}{2}$ & $+\tfrac{1}{2}$ & $\phantom{+}0$ & $0$ \\
    $\gen{Q}_-$ & $+\tfrac{1}{2}$ & $+\tfrac{1}{2}$ & $+\tfrac{1}{2}$ & $\phantom{+}0$ & $0$ \\
    $\gen{S}^+$ & $-\tfrac{1}{2}$ & $+\tfrac{1}{2}$ & $-\tfrac{1}{2}$ & $\phantom{+}0$ & $0$ \\
    $\gen{S}^-$ & $-\tfrac{1}{2}$ & $-\tfrac{1}{2}$ & $-\tfrac{1}{2}$ & $\phantom{+}0$ & $0$ \\
    \bottomrule
  \end{tabular}
\end{center}
We note in particular that $\gen{Q}_{\alpha}$ and $\gen{S}^{\alpha}$ preserve the length $\gen{L}$ while $\dot{\gen{Q}}$ and $\dot{\gen{S}}$ increase and decrease it by one, respectively. The last column in the table simply shows that that the algebra leaves the sector closed.

To leading order the generators acting on the fields take the simple form
\begin{equation}
  \begin{aligned}
    \gen{Q}_{\alpha} &= e^{+i\beta_0} \ketbra{\psi}{\phi_{\alpha}} , \\
    \gen{S}^{\alpha} &= e^{-i\beta_0} \ketbra{\phi^{\alpha}}{\psi} , \\
    \gen{D} &= \ketbra{\phi^{\alpha}}{\phi_{\alpha}} + \tfrac{3}{2} \ketbra{\psi}{\psi} , \\
    \gen{J}^{\alpha}{}_{\beta} &= \ketbra{\phi^{\alpha}}{\phi_{\beta}} - \tfrac{1}{2} \delta^{\alpha}{}_{\beta} \ketbra{\phi^{\gamma}}{\phi_{\gamma}} , \\
    \gen{J}_{\bullet} &= \tfrac{1}{2} \ketbra{\psi}{\psi} , \\
    \gen{L} &= \ketbra{\phi^{\alpha}}{\phi_{\alpha}} + \ketbra{\psi}{\psi} ,
  \end{aligned}
\end{equation}
with $\dot{\gen{Q}}$ and $\dot{\gen{S}}$ acting trivially (note that the above leads to $\gen{D} - \gen{J}_{\bullet} - \gen{L} = 0$). It is useful to introduce the Hamiltonian $\gen{H}$ and mass operator $\gen{M}$ which are defined such that\footnote{%
  The parameter $\beta_0$ corresponds to a similarity transformation $\gen{J} \to e^{+2i \delta\beta_0 \gen{J}_{\bullet}} \gen{J} e^{-2i \delta\beta_0 \gen{J}_{\bullet}}$
  under which $\beta_0 \to \beta_0 + \delta\beta_0$.
}%
\begin{equation}
  \begin{aligned}
    \acomm{\dot{\gen{Q}}}{\dot{\gen{S}}} &= \tfrac{1}{2} \bigl( \gen{H} + \gen{M} \bigr) , \\
    \acomm{\gen{Q}_2}{\gen{S}^2} &= \tfrac{1}{2} \bigl( \gen{H} - \gen{M} \bigr) .
  \end{aligned}
\end{equation}
To the leading order we then have
\begin{equation}
  \gen{H} = -\gen{M} = \tfrac{1}{2} ( \gen{D} + \gen{J}_{\bullet} ) - \gen{J}^1{}_1 = \ketbra{\phi^2}{\phi_2} + \ketbra{\psi}{\psi} ,
\end{equation}
so that the ground state of the form $\tr (\phi^1)^L$ is annihilated by the Hamiltonian, while the two excitations $\phi^2$ and $\psi$ have mass one.

We now want to find higher order corrections to this representation, by writing the generators as a series expansion in the coupling constant $h$
\begin{equation}
  \gen{J} = \gen{J}_{(0)} + h \, \gen{J}_{(1)} + h^2 \, \gen{J}_{(2)} + \dotsb ,
\end{equation}
imposing the commutation relations, the reality conditions and that the algebra preserves parity. The compact bosonic generators $\gen{J}^{\alpha}{}_{\beta}$, $\gen{L}$ and $\gen{J}_{\bullet}$ receive no corrections, but the supercharges as well as the dilatation operator $\gen{D}$ do. The latter we will express in terms of the spin-chain Hamiltonian $\gen{H}$. As we will see, the expansion of the charges of $\algSU(1|2)$ only contain terms that are of even order in the coupling constant, while the $\algSU(1|1)$ supercharges only come in at odd orders.

The first correction we find is a non-trivial contribution to the $\algSU(1|1)$ super charges
\begin{equation}
  \begin{aligned}
    \dot{\gen{Q}}_{(1)} &= + \tfrac{1}{2} \alpha_1 e^{-i(\beta_0 - \beta_1)} \epsilon_{\alpha\beta} \ketbra{\phi^{\alpha}\phi^{\beta}}{\psi} , \\
    \dot{\gen{S}}_{(1)} &= - \tfrac{1}{2} \alpha_1 e^{+i(\beta_0 - \beta_1)} \epsilon^{\alpha\beta} \ketbra{\psi}{\phi_{\alpha}\phi_{\beta}} .
  \end{aligned}
\end{equation}
The parameter $\beta_1$ corresponds to a similarity transformation using the operator $\gen{L}$, and $\alpha_1$ can be absorbed in a rescaling of the coupling constant $h$. 
At the second order we find corrections to the Hamiltonian
\begin{equation}
  \begin{aligned}
    \gen{H}_{(2)} &= \tfrac{1}{2}\alpha_1^2 \bigl( 
    \ketbra{\phi^{\alpha}\phi^{\beta}}{\phi_{\alpha}\phi_{\beta}} - \ketbra{\phi^{\beta}\phi^{\alpha}}{\phi_{\alpha}\phi_{\beta}} 
    + \ketbra{\phi^{\alpha}\psi}{\phi_{\alpha}\psi} - \ketbra{\psi\phi^{\alpha}}{\phi_{\alpha}\psi}
    \\ &\qquad\qquad
    + \ketbra{\psi \phi^{\alpha}}{\psi \phi_{\alpha}} - \ketbra{\phi^{\alpha}\psi}{\psi \phi_{\alpha}} 
    + 2 \ketbra{\psi\psi}{\psi\psi}
    \bigr) ,
  \end{aligned}
\end{equation}
and to the supercharges
\begin{equation}
  \begin{aligned}
    \gen{Q}_{\alpha}^{(2)} &= 
    \tfrac{1}{8} \alpha_1^2 e^{i\beta_0} \bigl(
    \ketbra{\phi^{\beta}\psi}{\phi_{\beta}\phi_{\alpha}} - \ketbra{\phi^{\beta}\psi}{\phi_{\alpha}\phi_{\beta}}
    + \ketbra{\psi\phi^{\beta}}{\phi_{\alpha}\phi_{\beta}} - \ketbra{\psi\phi^{\beta}}{\phi_{\beta}\phi_{\alpha}}
    \\ & \qquad\qquad
    + \ketbra{\psi\psi}{\phi_{\alpha}\psi} - \ketbra{\psi\psi}{\psi\phi_{\alpha}}
    \bigr)
    \\ &\quad +
    i \gamma_1 e^{i\beta_0} \bigl(
    \ketbra{\phi^{\beta}\psi}{\phi_{\beta}\phi_{\alpha}} + \ketbra{\psi\phi^{\beta}}{\phi_{\alpha}\phi_{\beta}}
    - \ketbra{\psi\psi}{\phi_{\alpha}\psi} + \ketbra{\psi\psi}{\psi\phi_{\alpha}}
    \bigr)
    \\ &\quad +
    i \gamma_2 e^{i\beta_0} \bigl(
\ketbra{\phi^{\beta}\psi}{\phi_{\alpha}\phi_{\beta}} + \ketbra{\psi\phi^{\beta}}{\phi_{\beta}\phi_{\alpha}}
    + \ketbra{\psi\psi}{\alpha_{\alpha}\psi} - \ketbra{\psi\psi}{\psi\phi_{\alpha}}
    \bigr)
    \\ &\quad +
    \tfrac{i}{2} \gamma_3 e^{i\beta_0} \bigl(
    \ketbra{\phi^{\beta}\psi}{\phi_{\beta}\phi_{\alpha}} + \ketbra{\psi\phi^{\beta}}{\phi_{\alpha}\phi_{\beta}}
    + \ketbra{\psi\psi}{\phi_{\alpha}\psi} - \ketbra{\psi\psi}{\psi\phi_{\alpha}}
    \bigr),
  \end{aligned}
\end{equation}
with $\gen{S}^{\alpha}_{(2)}$ given as the conjugate of the above expression. Shifts in the parameters $\gamma_1$, $\gamma_2$ and $\gamma_3$ are generated by similarity transformations with respect to the operators
\begin{equation}
  \begin{aligned}
    \gen{T}_1 &= \ketbra{\phi^{\alpha}\psi}{\phi_{\alpha}\psi} + \ketbra{\psi\phi^{\alpha}}{\psi\phi_{\alpha}} , 
    \\
    \gen{T}_2 &= \ketbra{\phi^{\alpha}\psi}{\psi\phi_{\alpha}} + \ketbra{\psi\phi^{\alpha}}{\phi_{\alpha}\psi} , 
    \\
    \gen{T}_3 &= \tfrac{1}{2} ( \ketbra{\phi^{\alpha}\psi}{\phi_{\alpha}\psi} + \ketbra{\psi\phi^{\alpha}}{\psi\phi_{\alpha}} ) + \ketbra{\psi\psi}{\psi\psi} .    
  \end{aligned}
\end{equation}
Note that $\gen{T}_3$ is an extension of $\gen{J}_{\bullet}$ to act on two sites and the term in $\gen{Q}_{\alpha}^{(2)}$ with coefficient $\gamma_3$ is proportional to $\gen{Q}_{\alpha}^{(0)}$.

It is straightforward to continue this expansion to higher orders in the coupling. However, let us now instead focus on how the eigenstates described by the above spin-chain Hamiltonian fall into representations of $\algPSU(1,1|2)_{\sL} \oplus \algPSU(1,1|2)_{\sR}$. To the zeroth order in the coupling, all states in the subsector are annihilated by the left-moving supercharges $\dot{\gen{Q}}$ and $\dot{\gen{S}}$. This means that they satisfy the 1/4-BPS condition $D_{\sL} = J_{\sL}$, and hence transform in a short representation of $\algPSU(1,1|2)_{\sL}$. When the coupling is turned on a generic state no longer saturates the 1/4-BPS bound, since the state receives an anomalous dimension. Hence, the state has to transform in a long $\algPSU(1,1|2)_{\sL}$ representation, and, indeed, is no longer annihilated by the $\algSU(1|1)_{\sL}$ generators $\dot{\gen{Q}}$ and $\dot{\gen{S}}$. Instead, these generators act on the state by \emph{adding or removing a spin-chain site}.

Here we have considered the $\algSU(1|1)_{\sL} \oplus \algSU(1|2)_{\sR}$ subsector of the massive spin chain. In this sector the Hamiltonian, as well as the full $\algSU(1|2)_{\sR}$ algebra, preserves the length of the spin chain. In a more generic sector, the left-moving and the right-moving supercharges and the Hamiltonian will all contain terms that relate states of different lengths.

A natural subsector for studying the massless modes is the 1/2-BPS sector which consists of the states that satisfy $D = J$. These are built up from the massive field $\phi^{++}$ and the 4+4 massless excitations $\chi_{\sL}^{+\pm}$, $\chi_{\sR}^{+\pm}$ and $T^{\pm\pm}$. The sector is preserved by the supercharges
\begin{equation}
  \gen{Q}^{\sL}_2 , \quad \gen{S}_{\sL}^2 , \quad \dot{\gen{Q}}_{\sL}^1 , \quad \dot{\gen{S}}^{\sL}_1 ,
  \qquad \text{and} \qquad
  \gen{Q}^{\sR}_2 , \quad \gen{S}_{\sR}^2 , \quad \dot{\gen{Q}}_{\sR}^1 , \quad \dot{\gen{S}}^{\sR}_1 ,
\end{equation}
and by the spin-chain Hamiltonian $\gen{H}$ and the mass operator $\gen{M}$. These charges generate the algebra $\algPSU(1|1)^4$, which is centrally extended by $\gen{H}$ and $\gen{M}$, which is the same symmetry algebra that was used in~\cite{Borsato:2014exa,Borsato:2016kbm} to determine the world-sheet S matrix. 

We will now show that the above generators do not have a well-defined expansion in the interaction length. For simplicity, let us consider only the three fields
\begin{equation}
  \phi \equiv \phi^{++} , \quad
  \chi \equiv \chi_{\sR}^{++} , \quad
  T \equiv T^{++} .
\end{equation}
Following the structure of the multiplets discussed in section~\ref{sec:massless-excitations}, we can write an ansatz for the first few orders of the expansions of the supercharges when acting on the above fields. For $\dot{\gen{Q}}_{\sR}^1$ and $\dot{\gen{S}}^{\sR}_1$ we have
\begin{equation}
  \begin{aligned}
    \dot{\gen{Q}}_{\sR}^2 &= \alpha_1 \ketbra{\chi}{T} 
    + g^2 \bigl( \delta_1 \ketbra{\chi\phi}{T\phi} + \delta_2 \ketbra{\phi\chi}{T\phi} + \delta_3 \ketbra{\chi\phi}{\phi T} + \delta_4 \ketbra{\phi\chi}{\phi T} \bigr)  
    + \dotsb , \\
    \dot{\gen{S}}^{\sR}_1 &= \bar{\alpha}_1 \ketbra{\chi}{T} 
    + g^2 \bigl( \bar{\delta}_1 \ketbra{T\phi}{\chi\phi} + \bar{\delta}_2 \ketbra{T\phi}{\phi\chi} + \bar{\delta}_3 \ketbra{\phi T}{\chi\phi} + \bar{\delta}_4 \ketbra{\phi T}{\phi\chi} \bigr) 
    + \dotsb ,
  \end{aligned}
\end{equation}
where we have introduced an  auxiliary coupling constant $g$ to keep track of the orders in the expansion. Similarly, the leading expansion of $\gen{Q}^{\sL}_2$ and $\gen{S}^{\sL}_2$ is given by
\begin{equation}
  \begin{aligned}
    \gen{Q}^{\sL}_2 &= g \bigl( \beta_1 \ketbra{\phi T}{\chi} + \beta_2 \ketbra{T \phi}{\chi} ) + \dotsb , \\
    \gen{S}_{\sL}^2 &= g \bigl( \bar{\beta}_1 \ketbra{\chi}{\phi T} + \bar{\beta}_2 \ketbra{\chi}{T \phi} \bigr) + \dotsb , 
  \end{aligned}
\end{equation}
These expressions are very similar to what we had in the massive sector discussed above. However, the massless excitations are all annihilated by the mass generator $\gen{M}$. Hence, the supercharges now satisfy the algebra
\begin{equation}
  \acomm{\gen{Q}^{\sL}_2}{\gen{S}_{\sL}^2} = \tfrac{1}{2} \gen{H} , \qquad
  \acomm{\dot{\gen{Q}}_{\sR}^1}{\dot{\gen{S}}^{\sR}_1} = \tfrac{1}{2} \gen{H} ,
\end{equation}
with exactly the same generator on the right hand side of both commutation relations. The expansion of the right-moving supercharges give a leading contribution to $\gen{H}$ of order $g^0$, which takes one field to one field. However, such a term can not appear from the anti-commutator of the two left-moving supercharges. Hence, we need to have $\alpha_1 = \bar{\alpha}_1 = 0$. Now, the left relation above starts at order $g^2$, while the right one starts at order $g^4$, which means that we also need to set the order $g$ coefficients of the left-moving supercharges to zero. Continuing in this fashion, we see order for order that there is no way to perturbatively deform the representations of the above algebra. Instead, the spin-chain Hamiltonian for the massless excitations is long-range even at the first non-trivial order.

A hint of the form of the Hamiltonian can be obtained from the massless dispersion relation
\begin{equation}
  E(p) = 2h\bigl|\sin\frac{p}{2}\bigr| .
\end{equation}
Let us consider a single massless excitation on a spin-chain of infinite length. The above dispersion relation can be written as a Fourier sum as
\begin{equation}
  E(p) = \frac{2h}{\pi} \sum_{n = -\infty}^{\infty} \frac{e^{ipn}}{n^2 - \tfrac{1}{4}} .
\end{equation}
The exponential in the sum can be interpreted as a hopping term in the Hamiltonian, where the excitation jumps $n$ sites. From this expression we see that the Hamiltonian involves interactions involving fields an arbitrary distance apart.

\renewcommand{\AdS}{AdS}
\renewcommand{\CFT}{CFT}
\renewcommand{\Sphere}{S}
\renewcommand{\Torus}{T}

\bibliographystyle{nb}
\bibliography{refs}

\end{document}